\documentclass{ceab}   

\usepackage{epsfig}     
\usepackage{graphicx}   

\usepackage{ceabbib}     
\usepackage[T1]{fontenc}
\usepackage{hyperref}
\usepackage{siunitx}  
\usepackage[dvipsnames]{xcolor}   

\def\EBV{\mbox{$E(4405-5495)$}}
\def\RV{\mbox{$R_{5495}$}}
\def\AV{\mbox{$A_{5495}$}}
\def\HII{\mbox{H\,{\sc ii}}}
\def\Teff{\mbox{$T_{\rm eff}$}}
\def\logg{\mbox{$\log g$}}
\def\logd{\mbox{$\log d$}}
\def\HII{H\,{\sc ii}}
\def\GG{\mbox{$G$}}
\def\GBP{\mbox{$G_{\rm BP}$}}
\def\GRP{\mbox{$G_{\rm RP}$}}

\begin{document}

\title{Extinction, the elephant in the room that hinders
       optical Galactic observations}
\def\gore{Extinction, the elephant in the room}
\def\even{Extinction, the elephant in the room}

\author{J. Ma\'{\i}z Apell\'aniz$^{1}$
\vspace{2mm}\\
\it $^1$Centro de Astrobiolog\'{\i}a, CSIC-INTA, campus ESAC.\\ 
\it Camino bajo del castillo s/n, \num[detect-all]{28692} Villanueva de la Ca\~nada, Madrid, Spain.
}

\maketitle

\begin{abstract}
Extinction is the elephant in the room that almost everyone tries to avoid when analyzing optical/IR data: astronomers
tend to find a quick fix for it that the referee will accept, but that does not mean such a solution is correct or even 
optimal. In this contribution I address three important issues related to extinction that are commonly ignored and 
present current and future solutions for them: [1] Extinction produces non-linear photometric effects, [2] the extinction
law changes between sightlines, and [3] not all families of extinction laws have the same accuracy.
\end{abstract}


\section{Introduction}

$\,\!$\indent Interstellar extinction has been studied for almost a century and for a novice in the topic it may be 
surprising how much was already known about it at the early stages of that period. \citet{Trum34} describes extinction 
phenomena in terms of monochromatic (for single lines), selective (color excess) and general absorption (or proper 
extinction if one includes scattering), and as the existence of obscuration effects (dark areas of the sky devoid of 
visible stars). \citet{BaadMink37} and 
\citet{Morg44} noticed that the extinction law towards Orion is peculiar, implying that a uniform extinction law does not
exist for the whole sky. \citet{StebWhit43} produced the first attempt to compute an extinction law as a function of 
wavelength.

The 1950s brought the advent of medium-scale (hundreds of stars) photoelectric $UBV$ surveys of OB stars and, from them, 
the use of the $Q$ parameter to measure their extinctions and estimate spectral types based on the behavior of the 
extinction law and the fact that the ratio of total to selective extinction $R_V\equiv A_V/E(B-V)$ is close to 3.0 for 
many sightlines \citep{JohnMorg53,HiltJohn56}. At around the same time, \citet{Blan56,Blan57} recognized the existence of
photometric bandwidth effects in the analysis of the photometry of those stars, leading to non-linear extinction
trajectories in color-magnitude and color-color diagrams and, eventually, to the dependence of $R_V$ not only on the type 
of dust but also on its amount as well as on the SED of the star. The following two decades saw a number of such optical 
extinction studies of OB stars using mostly the Johnson and Str\"omgren photometric systems.

The extension to other wavelength regimes started in the 1970s and took off with the launch of IUE in 1978 and the opening 
of moderately large ultraviolet samples and with the subsequent pioneering work of \citet{RiekLebo85} in the infrared. Those
led to the seminal work of \citet{Cardetal89}, the first family of UV-optical-IR extinction laws dependent on a
parameter, the ratio of total to selective extinction, to determine both the amount and type of dust along the sightline.

The most recent development in extinction studies has taken place in the last quarter of a century with the appearance of
large-scale photometric surveys, with sample increases of several orders of magnitude and the development of dedicated
pipelines for the uniform processing of the data. In the NIR the most important step was the publication of 2MASS in 2003
\citep{Skruetal06} and in the optical a number of surveys became available in the period such as SDSS \citep{Yorketal00} 
and IPHAS \citep{Drewetal05}. The final step was the publication of the different \textit{Gaia} data releases 
\citep{Browetal16}, with the availability of astrometry, (spectro)photometry, spectroscopy, and variability for 
$\sim 10^9$ stars. The combination of those surveys have brought us to a new era, both in terms of data volume and in 
the types of stars used to derive extinction, which are now of late type in most studies.

Why do I call extinction the elephant in the room? Because it is an obvious phenomenon in astronomy but it is undesired
and most researchers want to just correct for it in a simple and quick way. The object of many papers is to obtain
magnitudes and colors of astronomical objects and extinction is just a nuisance that gets in the way. Unfortunately, some
of the simple and quick solutions ignore the complexities of the problem and lead to biased results. In this contribution
I address the three main reasons why this happens:

\begin{itemize}
 \item[\textbf{1.}] \textbf{Extinction is a non-linear effect in magnitude or color.} In general, doubling the amount of 
       dust does not double $A_V$ or $E(B-V)$ and the same amount and type of dust does not produce the same $A_V$ or of 
       $E(B-V)$ in two stars with different SEDs. One needs to use monochromatic definitions for the amount and type of 
       dust, so neither $E(B-V)$ nor $R_V$ are appropriate. One should use their monochromatic equivalents, e.g.
       \EBV\ and \RV.
 \item[\textbf{2.}] \textbf{The extinction law changes between sightlines.} \RV\ is not the same everywhere. Many times it
       is within the canonical range of 3.0-3.2 but in a significant number of occasions it is outside that range (mostly
       $> 3.2$). \RV\ is not even constant within a small region, especially if it includes diverse ISM environments such
       as \HII\ regions. Furthermore, some measurements of \RV\ use invalid or outdated approximations.
 \item[\textbf{3.}] \textbf{No existing family of extinction laws accurately covers the UV-optical-IR range.} The two
       most commonly used, \citet{Cardetal89} and \citet{Fitz99} can have significant photometric residuals 
       already for stars with $\EBV\sim 1.0$. \citet{Maizetal14a} works well up to $\EBV\sim 2.0$ and does a reasonable
       job even around $\EBV\sim 3.0$ but does not extend to the UV and has the incorrect behavior in the IR.
\end{itemize}

In the next three sections I address each of those reasons and provide recipes when available. I end up analyzing prospects
for the future.

\section{Non-linearity of extinction}

$\,\!$\indent Extinction as a function of wavelength is defined as the ratio of the observed (extinguished) to the 
unextinguished spectral energy distributions (SEDs) expressed in magnitudes:

\begin{equation}
A(\lambda) = -2.5 \log_{10} \left(\frac{F_\lambda(\lambda)}{F_{\lambda,0}(\lambda)}\right).
\label{Alambda}
\end{equation}

The extinction in a given band such as $V$ is the difference between the observed and the unextinguished magnitudes, which
for a photon-counting detector such as a CCD can be expressed as:

\begin{equation}
A_V \equiv V - V_0 = -2.5 \log_{10} \left(\frac{\int P_V(\lambda)     F_\lambda(\lambda) \lambda\,d\lambda}
                                               {\int P_V(\lambda) F_{\lambda,0}(\lambda) \lambda\,d\lambda}\right),
\label{AV}
\end{equation}

\noindent where $P_V(\lambda)$ is the sensitivity curve for the $V$ band. In an equivalent manner, for the $B$ band we 
have:

\begin{equation}
A_B \equiv B - B_0 = -2.5 \log_{10} \left(\frac{\int P_B(\lambda)     F_\lambda(\lambda) \lambda\,d\lambda}
                                               {\int P_B(\lambda) F_{\lambda,0}(\lambda) \lambda\,d\lambda}\right).
\label{AB}
\end{equation}

To compute the values in Eqns~\ref{AV}~and~\ref{AB} we need to substitute $F_\lambda$ using Eqn.~\ref{Alambda} and define
$A(\lambda)$. The traditional (band-integrated) $B-V$ color excess is defined as the difference in extinction between the 
$B$ and $V$ bands:

\begin{equation}
E(B-V) \equiv A_B - A_V = (B-V) - (B-V)_0
\label{EBV}
\end{equation}

\noindent and the (band-integrated) ratio of total to selective extinction is defined as the ratio of 
Eqns.~\ref{AV}~and~\ref{EBV}:

\begin{equation}
R_V \equiv \frac{A_V}{E(B-V)}.
\label{RV}
\end{equation}

\begin{figure}[t]
 \centerline{
 \includegraphics[width=0.60\linewidth]{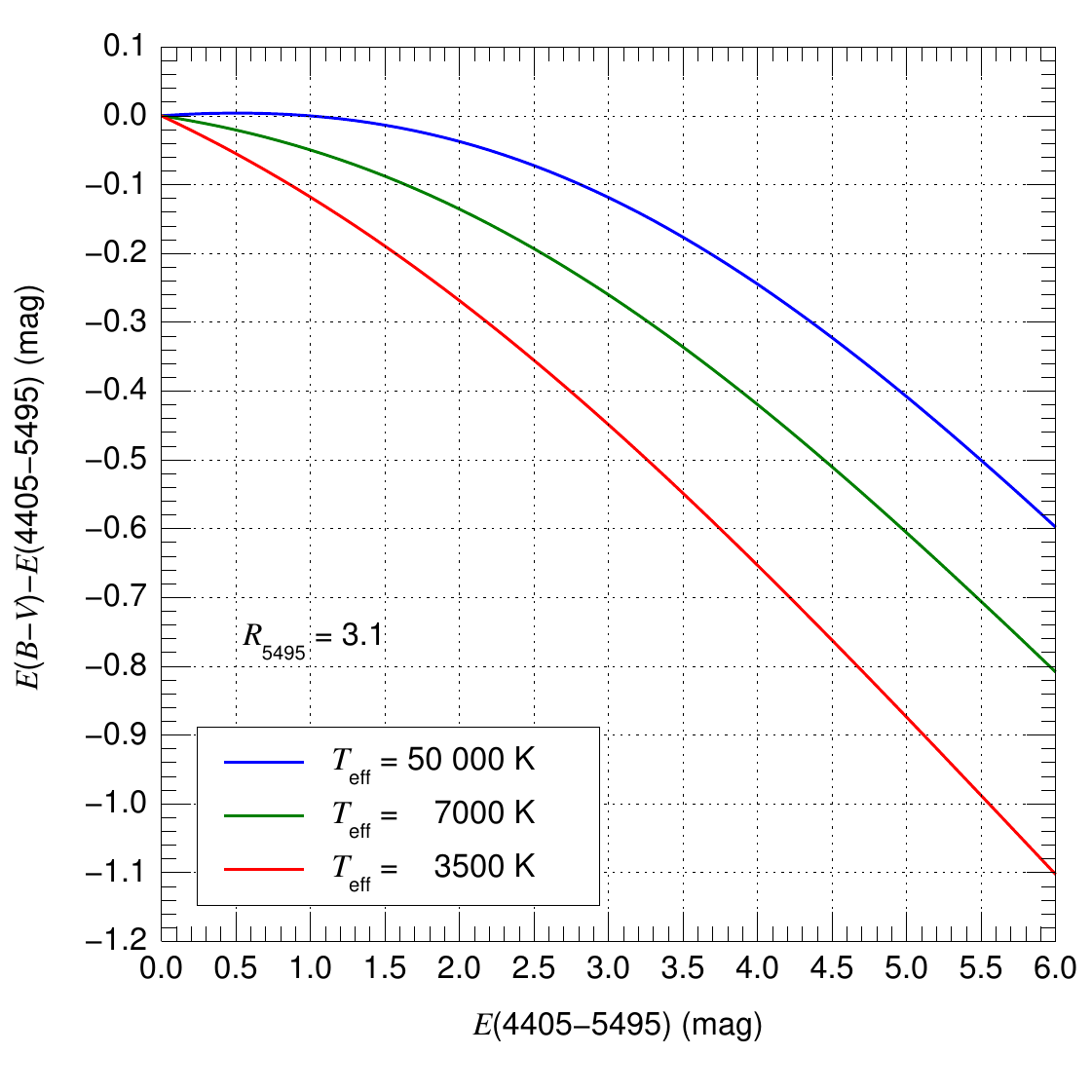}$\!\!\!\!$
 \includegraphics[width=0.60\linewidth]{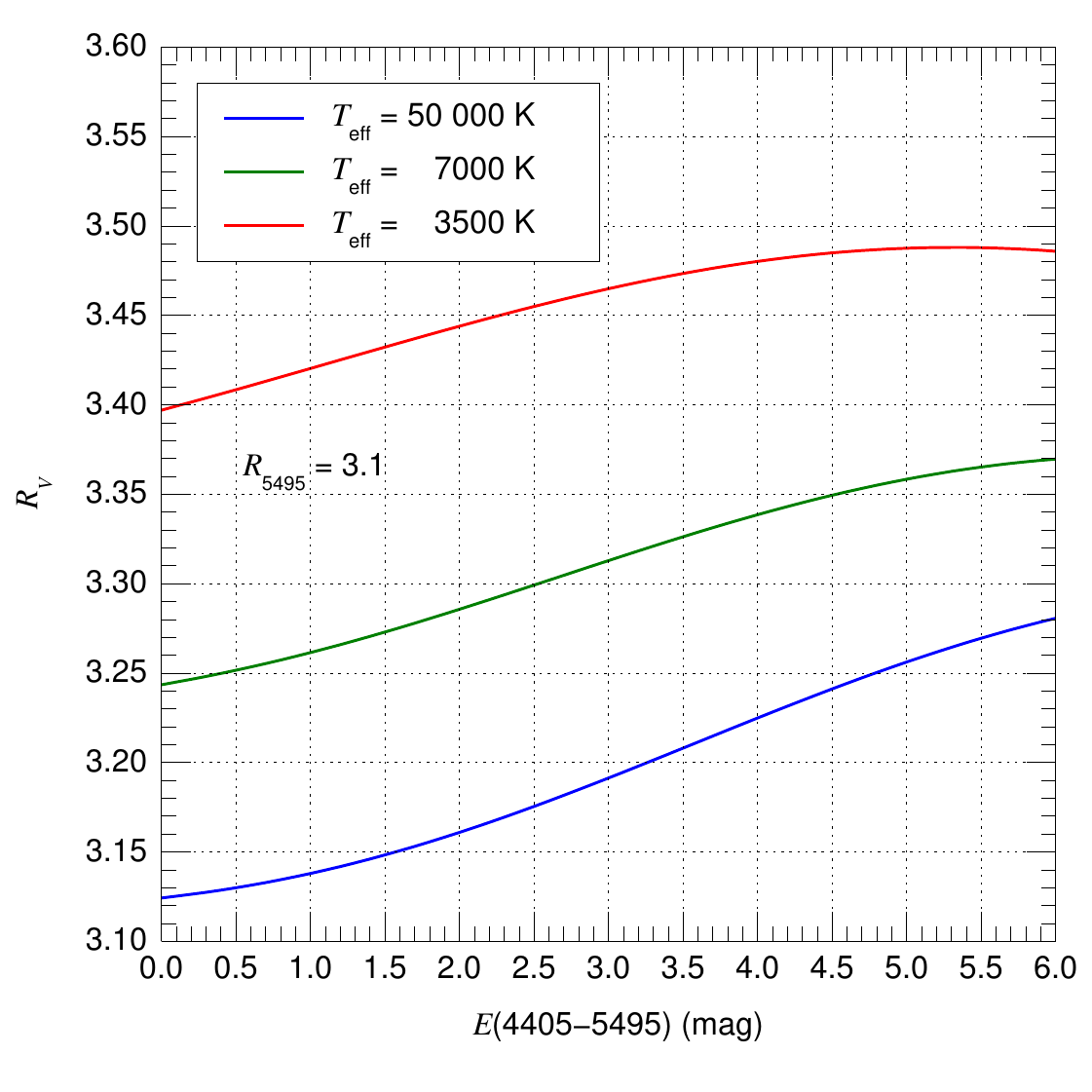}
 }
 \caption{(left) $E(B-V)-\EBV$ and (right) $R_V$ as a function of \EBV\ for a \citet{Maizetal14a} extinction law with 
          $\RV = 3.1$ and three main-sequence stars with different \Teff. $E(B-V) \approx \EBV$ and $R_V \approx \RV$ 
          only for hot stars with low extinctions.}
 \label{dEBVRV}
\end{figure}

The extinction law is simply the extinction normalized by some measurement of the amount of dust. This can be, for example,
the value of $A(\lambda)$ at a given wavelength or the difference between $A(\lambda)$ at two different wavelengths. What
should not be used as a normalization is a band-integrated extinction (such as those in Eqns.~\ref{AV}~or~\ref{AB}) or
color excess (such as that in Eqn.~\ref{EBV}). Why? \textbf{Because band-integrated quantities depend in a complex way on
the SEDs.} In other words, $A_V$ and $E(B-V)$ are not direct measurements of the amount of dust, as the same amount of dust
on the sightline of a red and a blue star yield different values of $A_V$ or $E(B-V)$. To show that, following
\citet{Maiz04c} I use as the amount of dust \EBV, the monochromatic equivalent of $E(B-V)$, and apply the family of
extinction laws of \citet{Maizetal14a}\footnote{The result does not change substantially for other families, see
\citet{Maiz13b} for the equivalent plots using \citet{Cardetal89}.} to different SEDs from the library of \citet{Maiz13a}.
Results are shown in Fig.~\ref{dEBVRV}, where it is clearly seen that the commonly used approximations of 
$E(B-V) \approx \EBV$ and $R_V \approx \RV$ are only valid for hot stars with low extinction. Indeed, the reason why in 
\citet{Maiz04c} the wavelengths of 4405~\AA\ and 5495~\AA\ were chosen as representative of $B$ and $V$, respectively, was
to allow for those approximations to be valid in that regime and, in that way, let the classic extinction measurements of
OB stars remain approximately valid.

\begin{figure}[t]
 \centerline{
 \includegraphics[width=1.20\linewidth]{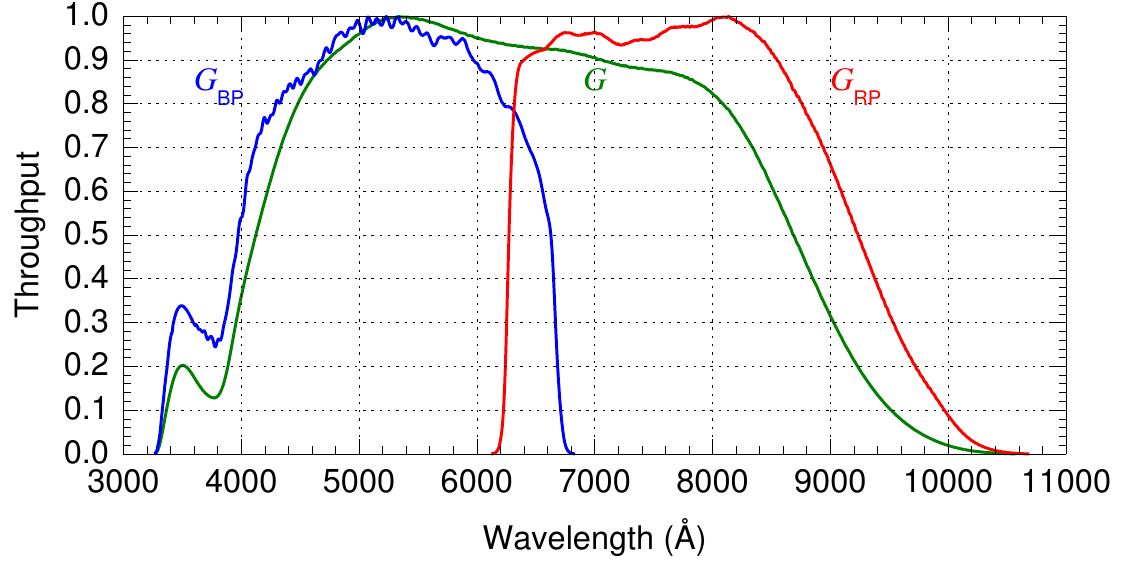}
 }
 \caption{Throughputs for the \textit{Gaia}~EDR3 \GG+\GBP+\GRP\ photometric system (Weiler et al. in prep.).}
 \label{Gaia_throughputs}
\end{figure}

As I have previously mentioned, bandwidth effects in photometry were known already in the 1950s. However, there is no
mention of them in \citet{Cardetal89}, where $E(B-V)$ is used as the amount of extinction and $R_V$ as the type of
extinction. \citet{Fitz99} mentions the difference between monochromatic and band-integrated quantities but he still uses
$E(B-V)$ and $R_V$ as parameters\footnote{The latter is called simply $R$ but its definition is that of $R_V$. Also, due to
the different definitions of central wavelengths, if $R$ is considered as the monochromatic definition for the extinction
law, then it is not exactly \RV.}. For their
samples the difference is small, as both papers use OB stars with low extinction (see below). Still, the formulae in those 
papers are inconsistent and need to be adapted to monochromatic quantities when implemented to allow them to be applied to 
SEDs of arbitrary temperatures and extinctions.

Despite claims to the contrary, bandwidth effects are always present. If they have been ignored in the past it may have been
because samples were more restricted in temperature and extinction and because past photometric calibrations were not as 
accurate as current ones (see \citealt{Maiz05b,Maiz06a,Maiz07a} for examples). Having said that, non-linearity starts 
becoming noticeable at lower extinctions for broader filters than for narrower ones (e.g. Johnson vs. Str\"omgren) and for 
wavelength regions where the extinction law is steeper. Those conditions make non-linearity a stronger effect for the
\textit{Gaia} \GG+\GBP+\GRP\ system than for other optical photometric systems, as the \GG\ band is very broad 
(Fig.~\ref{Gaia_throughputs}) and the photometric callibration of \textit{Gaia} is better than that of any optical 
ground-based system \citep{MaizWeil18}.

\begin{figure}[ht!]
 \centerline{
 \includegraphics[width=1.20\linewidth]{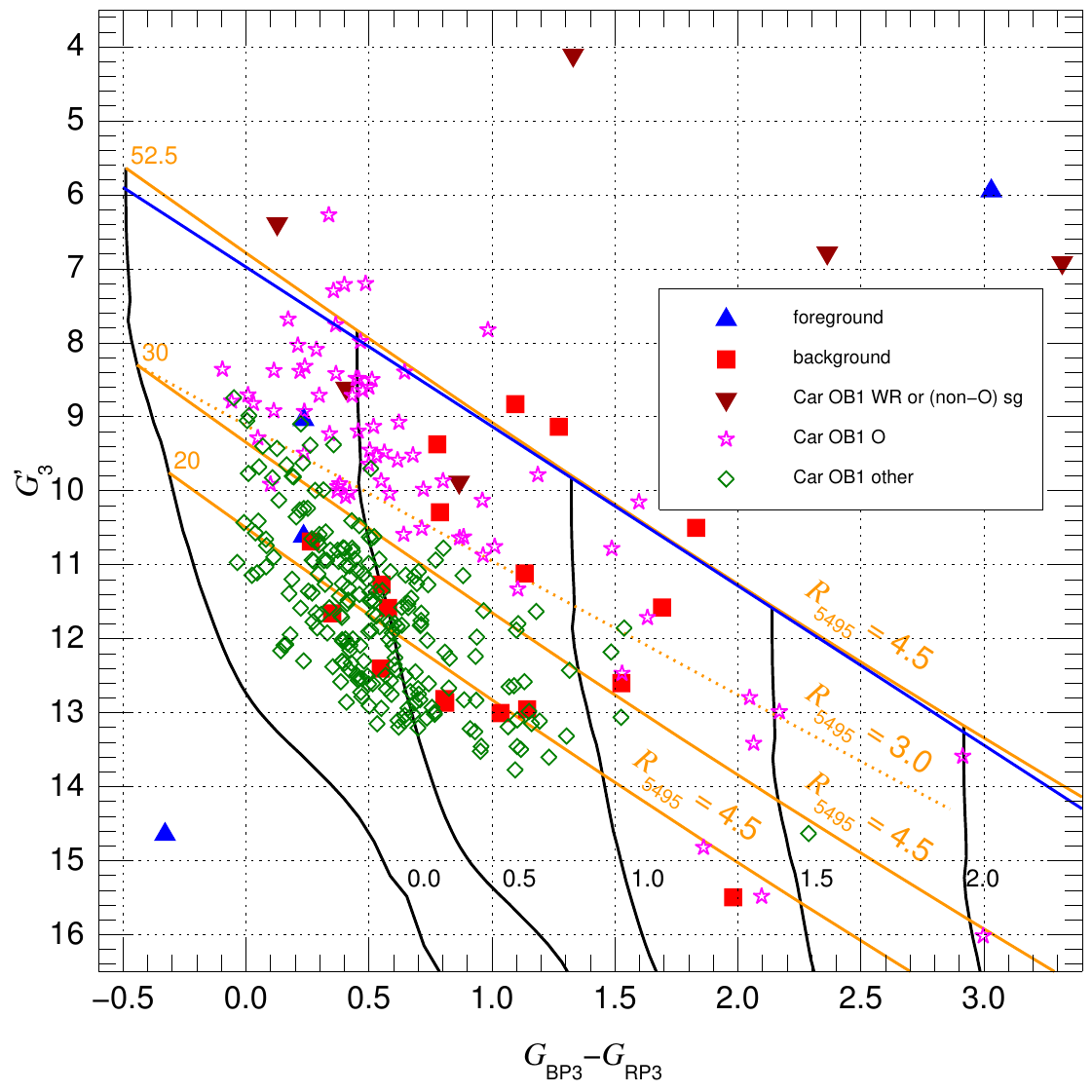}
 }
 \caption{\textit{Gaia}~EDR3 Carina~OB1 CMD adapted from \citet{Berletal23}. The orange lines show extinction tracks for MS 
          stars of different \Teff\ (indicated in kK at their left ends) and \RV\ (also labelled). For the 52.5~kK 
          extinction track with $\RV = 4.5$ I have also drawn a blue straight line next to it to show the curvature effect.}
 \label{Gaia_CMD}
\end{figure}

\begin{figure}[ht!]
 \centerline{
 \includegraphics[width=1.20\linewidth]{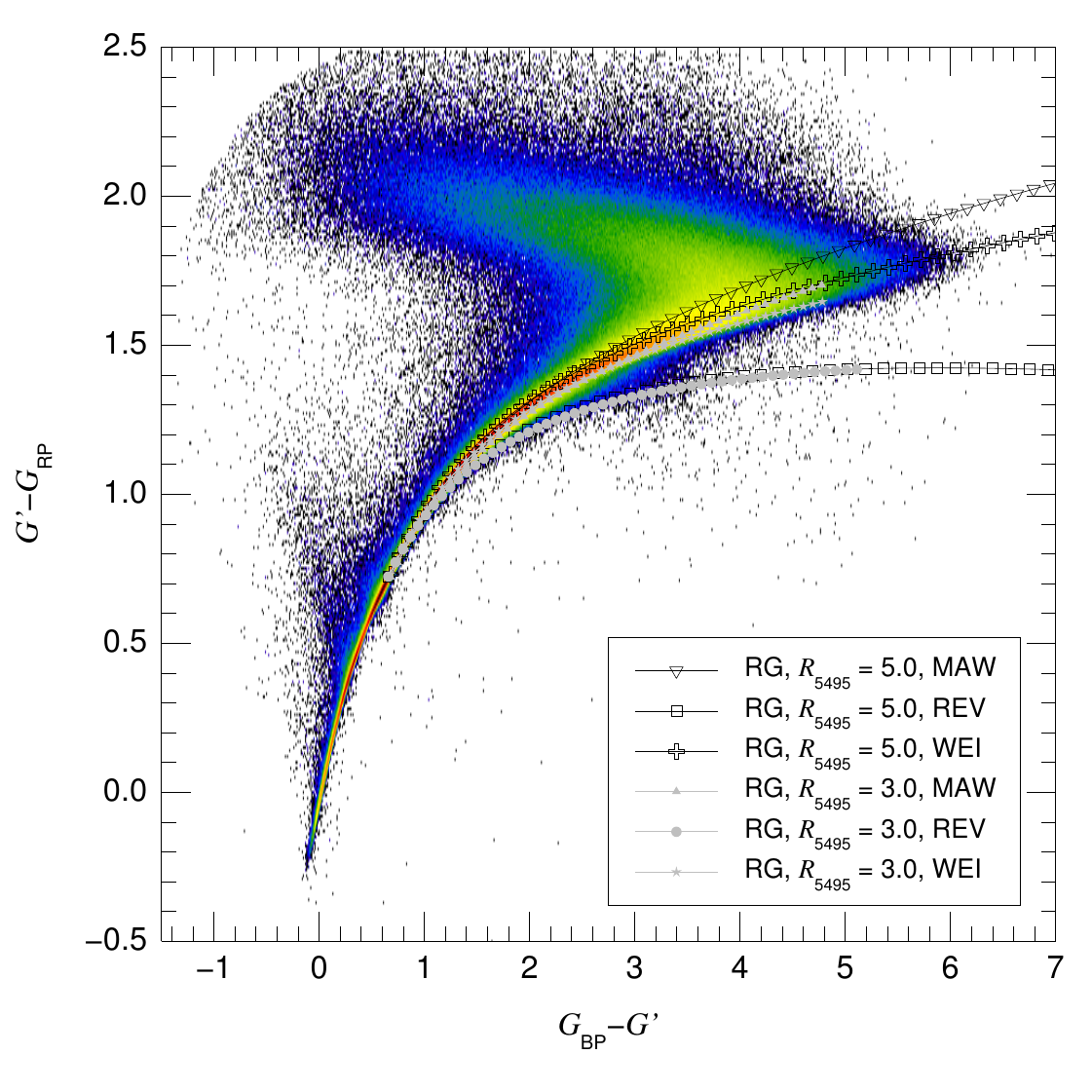}
 }
 \caption{\textit{Gaia}~DR2 color-color diagram from \citet{MaizWeil18}. The lines show the extinction tracks for
          red-clump stars with different assumptions about \RV\ and the filter throughputs (the purpose of that paper was to
          identify those), with the associated symbols plotted at intervals of \EBV\ of 0.1~mag.}
 \label{Gaia_BG_GR}
\end{figure}

\textit{Gaia} data processing correctly differentiates between monochromatic and band-integrated quantities. For example,
Apsis \citep{Creeetal23a} provides one monochromatic extinction ($A_0$, with the wavelength being 5414~\AA), three 
band-integrated extinctions ($A_G$, $A_{\rm BP}$, and $A_{\rm RP}$), and one band-integrated color excess or reddening 
[$E(\GBP-\GRP)$], each one adequately identified. Non-linearity appears both in \textit{Gaia} CMDs and color-color diagrams.
A sample CMD is shown in Fig.~\ref{Gaia_CMD}, adapted from the analysis of Carina~OB1 by \citet{Berletal23}. The extinction
tracks are shown in orange and computed for different \Teff\ and \RV\ values using the family of \citet{Maizetal14a}. As the
vertical scale comprises 13 magnitudes, the curvature is not apparent at first but plotting a straight line next to one of
them (in blue) shows that the effect amounts to several tenths of a magnitude over the extinction range plotted in this case
(most stars that appear in the plot are blue MS stars that have been displaced towards the lower right corner by the strong
differential extinction in the region). A sample color-color diagram is shown in Fig.~\ref{Gaia_BG_GR}, as it appeared in
\citet{MaizWeil18}. The stellar locus is highly curved due to extinction and the different values of \Teff, with both effects
being relatively similar due to the \textit{Gaia} filter design, and follows the result of the throughputs in that paper
(labelled as MAW) with $\RV = 3.0$ quite well (the extension towards the upper left part of the diagram is caused mostly by 
stars with contaminated photometry due to crowding or nebulosity). 

\begin{figure}[ht!]
 \centerline{
 \includegraphics[width=1.20\linewidth]{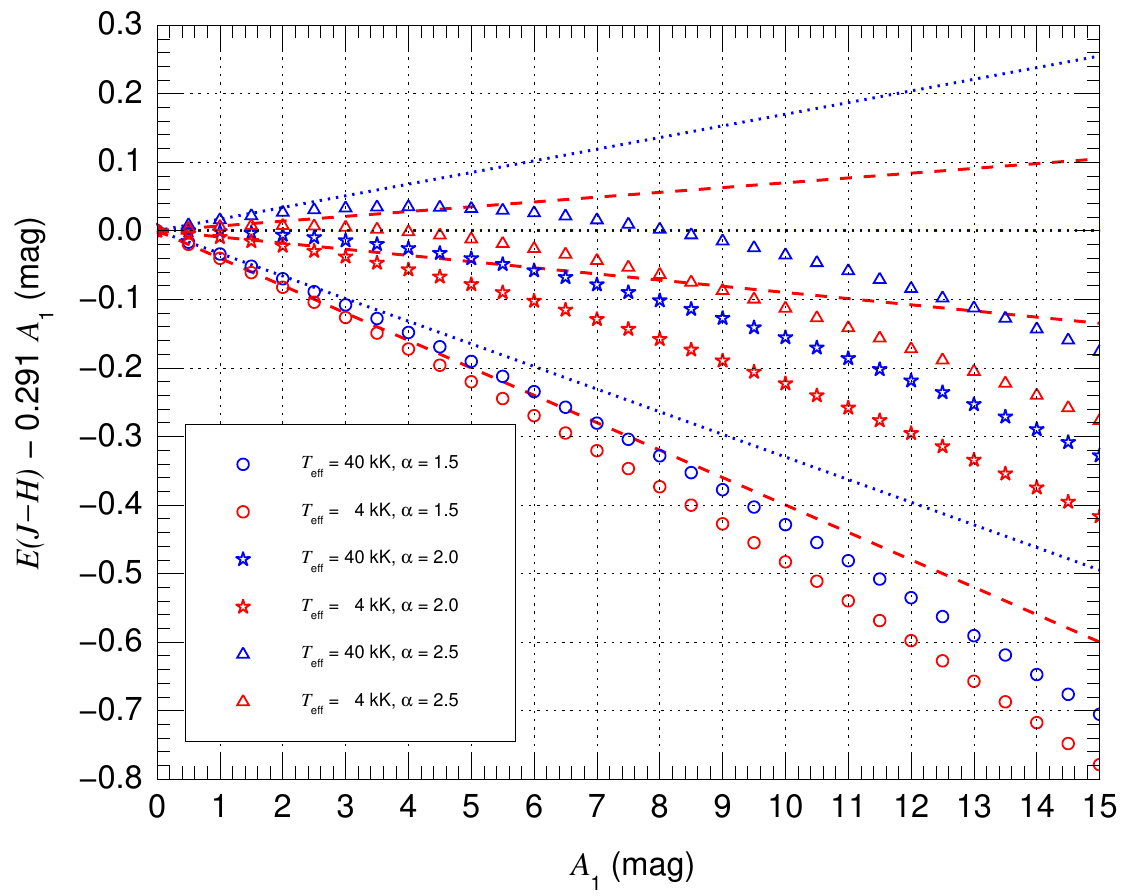}
 }
 \caption{Band-integrated $E(J-H)$ color excess as a function of the monochromatic amount of extinction $A_1$ from
          \citet{Maizetal20a}. Two effective \Teff\ (4~kK and 40~kK) and three values of the NIR slope $\alpha$ (1.5, 2.0, 
          and 2.5) are used. Symbols are used for the actual values and lines for the extrapolation from the low-extinction 
          regime.}
 \label{EJH}
\end{figure}

In the NIR, extinction is smaller than in the optical for the same dust column but, at the same time, we can reach higher
extinction values and non-linear effects appear sooner or later. As discussed in \citet{Maizetal20a}, if non-linearity is
ignored in IR studies, extinction measurement can be compromised. Some papers, such as
\citet{SteaHoar09,WangChen19,Noguetal21b} correctly deal with it but others do not. I show this in Fig.~\ref{EJH}, which is
taken from the Appendix in \citet{Maizetal20a}. As we increase $A_1$, the monochromatic extinction at 1~$\mu$m, the color
excess $E(J-H)$ clearly deviates from a non-linear behavior. The effect is stronger for higher values of $\alpha$, the
exponent in the assumed power-law NIR extinction. As we will see later, the current preferred value is
$\alpha\sim 2.2$, so the effect is easy to notice already at relative lower values of $A_1$.

As a summary of this section, extinction has a clear non-linear behavior in CMDs and color-color diagrams. The corollary is
that one should not apply the commonly used approximation of defining a central wavelength for a filter and extinguish the
photometry assuming that only the extinction law at that point is relevant. Extinction should be calculated using
integrals such as Eqns.~\ref{AV}~and~\ref{AB}. As that may be computationally too expensive when dealing with large samples
and fitting procedures, one alternative is to compute a grid for the desired values of the extinction and interpolate with
splines within the grid. This is the approach used by CHORIZOS \citep{Maiz04c}.

\section{Variability of the extinction law}

\subsection{Introduction}

$\,\!$\indent As I mentioned in the introduction, the variability of the extinction law in the optical/IR has been known 
since the 1930s but it was not until the 1980s that the much stronger variability in the UV was discovered with IUE.
\citet{Seat79} produced an average UV extinction law from OAO-2, Copernicus, and TD-1 data but subsequent work revealed
variability among Galactic sightlines \citep{Hechetal82,Massetal83,CardSava88} and between those and LMC
\citep{Howa83,FitzSava84,ClayMart85,Fitz85,Fitz86} and SMC \citep{Roccetal81,Lequetal82,Prevetal84,GordClay98,MaizRubi12}
sightlines.

The studies of extinction of the 1980s culminated in the work of \citet{Cardetal89}, who, as previously mentioned, produced
the first \RV-dependent family of extinction laws. That family has a uniform behavior in the IR (based on
\citealt{RiekLebo85}) and a highly variable dependence on \RV\ as a single parameter in the optical and UV. Another commonly 
used family of extinction laws, that of \citet{Fitz99}, appeared a decade later and has a similar behavior to the
\citet{Cardetal89} one: non-variable in the IR and dependent on an $R$-like parameter in the optical and UV. In the next
section I analyze their similarities and differences in more detail. I also address extinction-law variability in the IR
later on. For the time being, I just point out that, given the range of extinctions covered by the previously mentioned two
families of extinction laws, such IR extinction variability would have been very difficult to measure with the data 
available in the 1980s and 1990s.

\subsection{Methodology} 

$\,\!$\indent Before addressing the variability of the extinction law, I need to discuss how the extinction law and 
extinction in general are measured. In principle, $A(\lambda)$ is determined by Eqn.~\ref{Alambda}, so what we need is the 
observed SED determined from spectrophotometry, divide it by the unextinguished SED, obtain the logarithm, and multiply the 
result by $-2.5$. As expected, that is more easily said than done for two reasons: [1] We do not know a priori the 
unextinguished SED and [2] in most cases we do not have spectrophotometric data but only photometry. I address those two 
issues starting with the first one.

\textbf{The unextinguished SED.} The traditional method used, for example, in most of the 1980s work with IUE data, is 
called the pair method. One obtains
the spectrum of an identical star with no extinction and uses it as a reference or standard for the unextinguished SED. The 
three problems associated with the pair method are: obtaining the ratio of the distances needed to normalize the fluxes, 
determining that the two stars are indeed identical, and finding the reference star itself. Of those, the third one is the 
most difficult one: OB stars are concentrated in the plane of the Milky Way and are usually at considerably distances, so 
extinction is almost never negligible \citep{MaizBarb18}. Due to those problems, in the last two decades most works have 
shifted towards using synthetic SEDs as a reference instead of observed ones \citep{Maiz04c,FitzMass05b}. For this
alternative method we still have the first problem (flux normalization due to distance) and the last two are substituted by
another two: making sure that the grid of synthetic SEDs is spectrophotometrically accurate and determining the parameters
of the star (\Teff\ being the most important one but also \logg\ and metallicity in some cases) to select the specific SED
in the grid. I discuss these issues below. 

\textbf{Photometry instead of spectrophotometry.} Analyses of UV extinction using IUE data in the 1980s had access to
spectrophotometry in the ultraviolet but even those complemented the optical and IR ranges with photometry. As most current
works use large-scale photometric optical/IR surveys as their data source for extinction determinations, photometry has
become the most popular method to determine extinctions. Photometry has the advantage of large sample sizes but introduces
problems with calibration for both the analyzed stars and for the standard system (see below) and only samples the 
extinction law at a coarse wavelength grid which then needs to be interpolated or modelled.

There are actually two related problems we are dealing with. One is determining the family of extinction laws that applies
to an astrophysical regime (such as a given galaxy or environment) and another one calculating the extinction for a set of
stars. In this section I start with the second problem, which assumes we already know the family of extinction laws, and in
the next section I will address the more complex problem of determining the family of extinction laws. To determine the
extinction parameters (amount and type) we need to fit the observed photometry to the synthetic photometry derived from the 
SED models. Using $\chi^2$ minimization, the problem can be expressed with these three equations:

\begin{equation}
\chi_{\rm red}^2 = \frac{1}{N-{\rm DoF}} = 
\sum_{i=1}^{N} \frac{(\fcolorbox{red}{white}{$m_{i,{\rm obs}}$}-m_{i,{\rm syn}})^2}
                    {\fcolorbox{red}{white}{$\sigma_{i,{\rm obs}}^2$}} 
\fcolorbox{violet}{white}{$\sim 1$}
\label{chired}
\end{equation}

\begin{equation}
m_{i,{\rm syn}} = -2.5 \log_{10} 
  \left(\frac{\int \fcolorbox{red}{white}{$P_i(\lambda)$}\,                                   F_\lambda(\lambda)  \,\lambda\,d\lambda}
             {\int \fcolorbox{red}{white}{$P_i(\lambda)$}\,\fcolorbox{OliveGreen}{white}{$F_{\lambda,r}(\lambda)$}\,\lambda\,d\lambda}\right) 
+ \fcolorbox{OliveGreen}{white}{${\rm ZP}_{i,r}$}
\label{misyn}
\end{equation}

\begin{equation}
F_\lambda(\lambda) = \fcolorbox{blue}{white}{$F_{\lambda,0}(\lambda)$}\,
 e^{-0.4\fcolorbox{orange}{white}{\scriptsize$A(\lambda)$}}
\label{Flambda}
\end{equation}

\noindent with the following definitions:

\begin{itemize}
\addtolength{\itemsep}{-7pt}
 \item $\chi_{\rm red}^2$: Goodness of fit (reduced chi square).
 \item $N$: Number of filters/passbands.
 \item DoF: Degrees of freedom of fit (number of parameters).
 \item $m_{i,{\rm obs}}$: Observed magnitude for filter/passband $i$.
 \item $m_{i,{\rm syn}}$: Synthetic magnitude for filter/passband $i$.
 \item $\sigma_{i,{\rm obs}}$: Observed magnitude uncertainty for filter/passband $i$.
 \item $P_i(\lambda)$: Sensitivity curve/throughput for filter/passband $i$. 
 \item $F_\lambda(\lambda)$: Observed SED.
 \item $F_{\lambda,r}(\lambda)$: Reference system SED ($r$ = Vega, ABmag, STmag).
 \item ZP$_{i,r}$: Photometric zero point for filter $i$ in the reference system $r$.
 \item $F_{\lambda,0}(\lambda)$: Unextinguished SED.
 \item $A(\lambda)$: Extinction as a function of $\lambda$ in mag.
\addtolength{\itemsep}{7pt}
\end{itemize}

The logical procedure is to generate an extinguished grid of SEDs using Eqn.~\ref{Flambda} as a function of intrinsic
(such as \Teff\ and \logg\ or luminosity class) and extrinsic (such as amount and type of dust) parameters, including
distance if it is not independently determined. The extinguished SED grid is transformed into an extinguished magnitude grid
using Eqn.~\ref{misyn} and the parameters and their uncertainties are obtained through the fitting process that uses 
Eqn.~\ref{chired}. This is the procedure used in CHORIZOS \citep{Maiz04c} and in other equivalent codes. However,
the devil is in the details and there are five different aspects that have to be considered. I describe them here, each
color-coded in Eqns.~\ref{chired}~to~\ref{Flambda} above.

\textbf{\textcolor{red}{Photometric calibration.}} There are three related aspects of photometric calibration to be
considered: the absence of biases in the observed magnitudes, the proper estimation of the photometric uncertainties, and 
the correct characterization of the sensitivity curves. As already mentioned, old-style ground-based systems such as Johnson
and Str\"omgren have problems inherent with the small samples analyzed by each author and with the presence of 
inconsistencies among different authors, leading to relatively large realistic uncertainties \citep{Maiz06a,Maiz07a}.
\textit{Gaia} photometry, on the other hand, has the opposite problem: the large sample and consistent reduction leads to
calibrations that are so accurate that small systematic effects can be detected, thus leading to improvements in the
calibration itself \citep{MaizWeil18}. In this respect, it is important to distinguish that there are two types of
photometric uncertainties. Internal uncertainties are those that do not make a comparison with an external system and, as
such, are appropriate for variability studies such as in \citet{Maizetal23}. However, if we want to compare our observed
magnitudes with synthetic ones, we also have to include the uncertainties in the knowledge of the photometric system
(external uncertainties). The easiest thing to do is to use a calculated minimum uncertainty or to add a systematic 
uncertainty in quadrature to the observed one \citep{MaizWeil18}. If this is not done, chances are that we will have
problems in our final check (see below). A similar issue can appear if our source is intrinsically variable and our 
photometry is obtained in different epochs, which is why a check for source variability is convenient \citep{Maizetal23}.

\textbf{\textcolor{OliveGreen}{Standard calibration.}} Photometric systems are calibrated with a reference SED that can be
an astrophysical source such as Vega (Vega system) or a simple function such as in the ABmag or STmag systems (see e.g.
\citealt{synphot}). In the case of an astrophysical source, it needs to be determined independently 
\citep{BohlGill04a,Bohl07,Bohl14}. In all cases, the photometric zero point (in some references, called the differential
zero point) needs to be calculated. It is a quantity that is supposed to be close to zero but seldom is actually zero due to
the way photometric surveys are calibrated (see \citealt{Maiz07a,MaizWeil18} and references therein).

\textbf{\textcolor{blue}{Model SEDs.}} Nowadays there are a number of choices for SED grids for stellar atmospheres of
different characteristics. However, before using them one needs to read the fine print, as some are optimized to accurately
describe stellar lines but not so much the absolute fluxes or have a detailed analysis in just some wavelength regions. One 
way to test them is to compare the derived synthetic photometry derived with the observed magnitudes of low-extinction 
stars. As an example, CHORIZOS uses a tailored SED grid built from different sources \citep{Maiz13a}. For hot 
stars, the SEDs are mostly from TLUSTY \citep{LanzHube03,LanzHube07} and accurately represent the optical colors of O and B 
stars. However, the NIR synthetic colors derived from TLUSTY are incorrect, so for long wavelengths the \cite{Munaetal05} 
grid is used instead. Also, one should always check whether the grid includes the type of system we are observing: stars 
with emission lines or IR excesses and binary systems with components of very different \Teff\ will usually need to be 
analyzed with an SED grid with additional parameters.

\textbf{\textcolor{orange}{Extinction.}} The main issue here is extinguishing the model SEDs (Eqn.~\ref{Flambda}) before
integrating in wavelength (Eqn.~\ref{misyn}) to correctly account for non-linear photometric effects, as discussed in the
previous section. Furthermore, some papers use simple linear approximations to determine the amount and type of extinction
from color-color relations: those determinations of $E(B-V)$ and $R_V$ should be distrusted in general (and, once again, one
should use monochromatic quantities for the amount and type of extinction, not band-integrated ones). 

\textbf{\textcolor{violet}{Final check.}} The last step to perform is to verify the goodness of the fit. Fitting a 
functional form to a data set and obtaining a result does not mean that the result makes sense: a straight line can be fit
to data extracted from a parabola to derive a slope and an intercept but that does not mean that those two
quantities accurately describe the data. Unfortunately, this step is omitted in many papers and one is left to wonder
whether the results make sense or not. Going over the four previous aspects, if the fit is not as good as it should it could
be due to a number of reasons: [1] The observed magnitudes may have systematic effects, their random uncertainties may be
underestimated, or the sensitivity curve may not be properly characterized. [2] The photometric zero points may be incorrect.
[3] The model SEDs may not properly represent the object. [4] The extinction law may be incorrect. Any of these reasons may
apply in different situations. Sometimes the problems will be in the calibration, in others we may have an interesting
astrophysical case, such a star with a previously undetected IR excess. One way to understand what is going on is to analyze
the fit residuals, as those may point out to e.g. systematic effects as a function of the amount of extinction, a telltale
sign of issues with the extinction law (see \citealt{Maizetal14a} for an example).

\subsection{Results}

\begin{figure}[hpt!]
 \centerline{
 \includegraphics[width=0.98\linewidth]{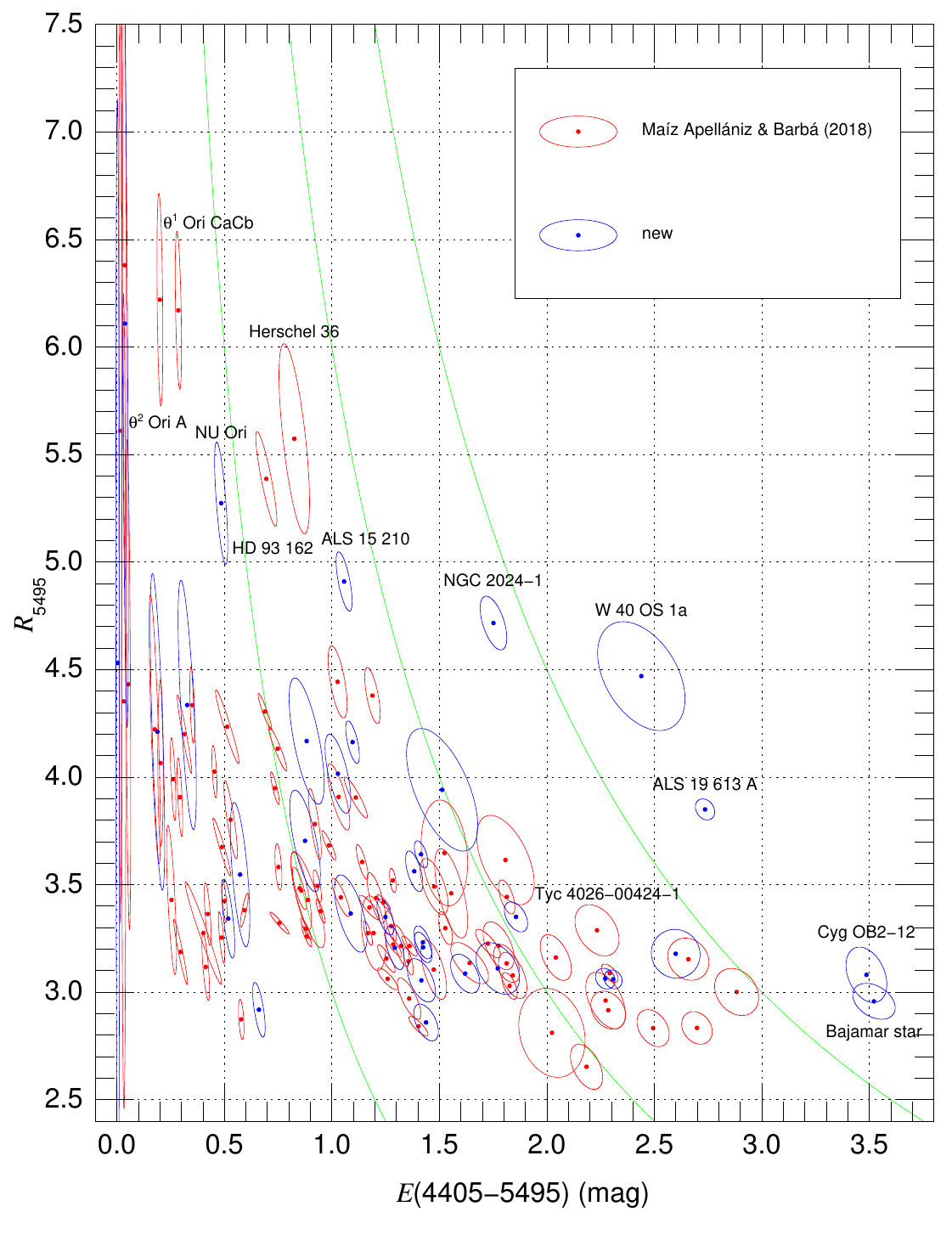}
 }
 \caption{\RV\ as a function of \EBV\ for the sample of 120 OBA stars in \citet{Maizetal21a}. The color coding indicates 
          whether the data appeared in \citet{MaizBarb18} or not. Green lines correspond (from left to right) to the values 
          of \AV\ of 3,~6,~and~9~mag.}
 \label{EBV_RV}
\end{figure}

$\,\!$\indent In \citet{MaizBarb18} we used the method described above to derive \EBV\ and \RV\ for 562 Galactic O stars. In
a subsequent paper \citep{Maizetal21a}, we selected a subsample of those and added some OB stars with notoriously high \EBV\
or \RV\ to obtain a golden sample of 120 OBA stars with measurements not only of those extinction properties but also of 
interstellar K\,{\sc i} and C$_2$. We also detected a very broad interstellar absorption band of unknown origin centered 
around 7700~\AA\ that should be incorporated into future famiies of extinction laws. 

The results for the golden sample are shown in Fig.~\ref{EBV_RV} and are derived from a combination of NIR (mostly 2MASS) 
and optical photometry (\textit{Gaia}, Tycho-2, Johnson, and Str\"omgren). In general, it is difficult to determine \RV\ from
optical photometry alone, as the information encoded at longer wavelengths is needed to determine \AV. The most important
result from Fig.~\ref{EBV_RV} is that there is a large scatter in the Galactic \RV\ values: from a floor around 2.6 to 
values higher than 6.0. A similar range (3.1-6.7) was found for the 30~Dor sample in \citet{Maizetal14a} using HST
photometry. The distributions in \RV\ and \EBV\ are not independent: the scatter in \RV\ decreases as \EBV\ increases and,
in the limit of high extinction, most \RV\ values concentrate in the 2.8-3.2 range. Therefore, the canonical values of
3.0-3.2 are still valid, but only as an approximation in that circumstance.

\begin{figure}[t]
 \centerline{
 \includegraphics[width=0.60\linewidth]{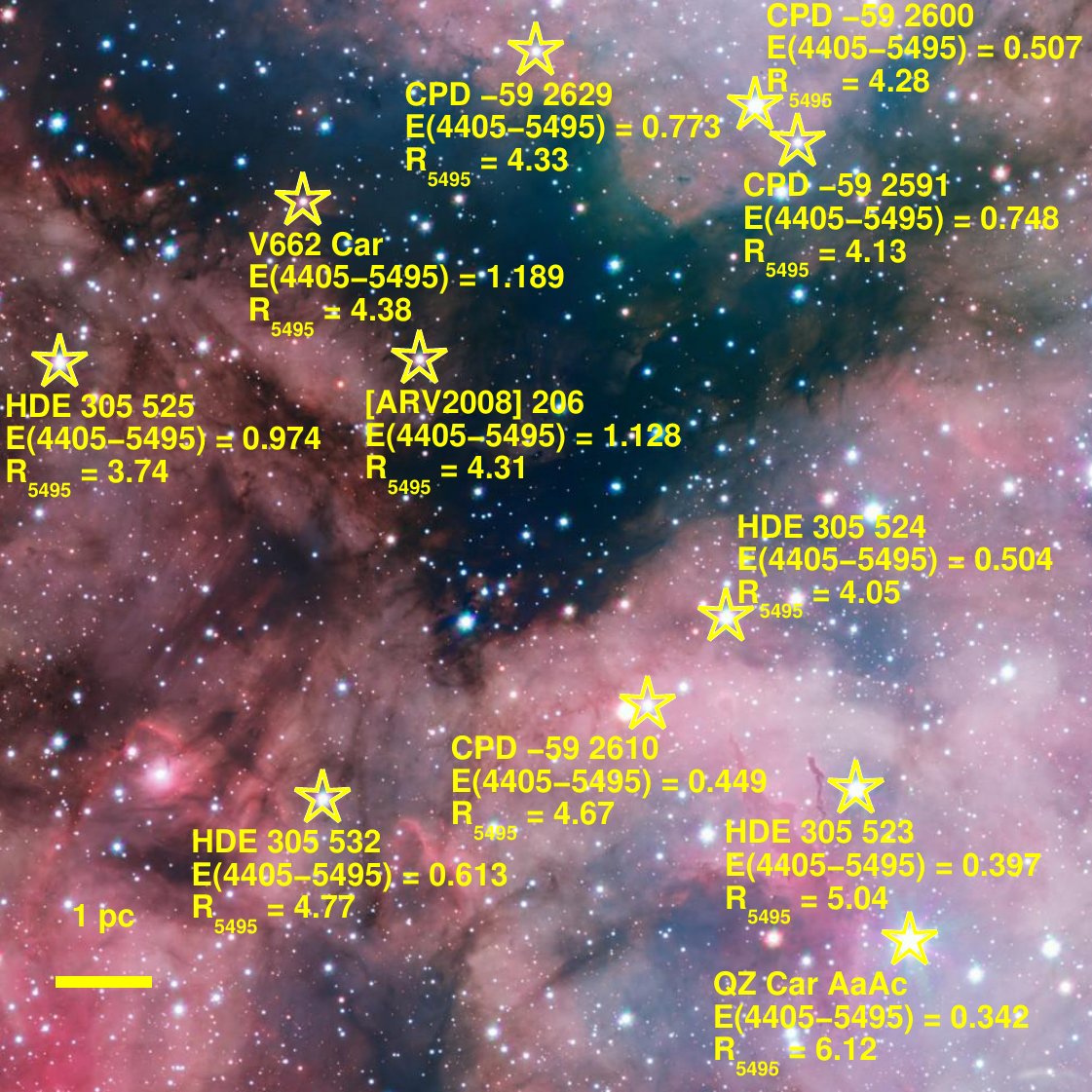} \
 \includegraphics[width=0.60\linewidth]{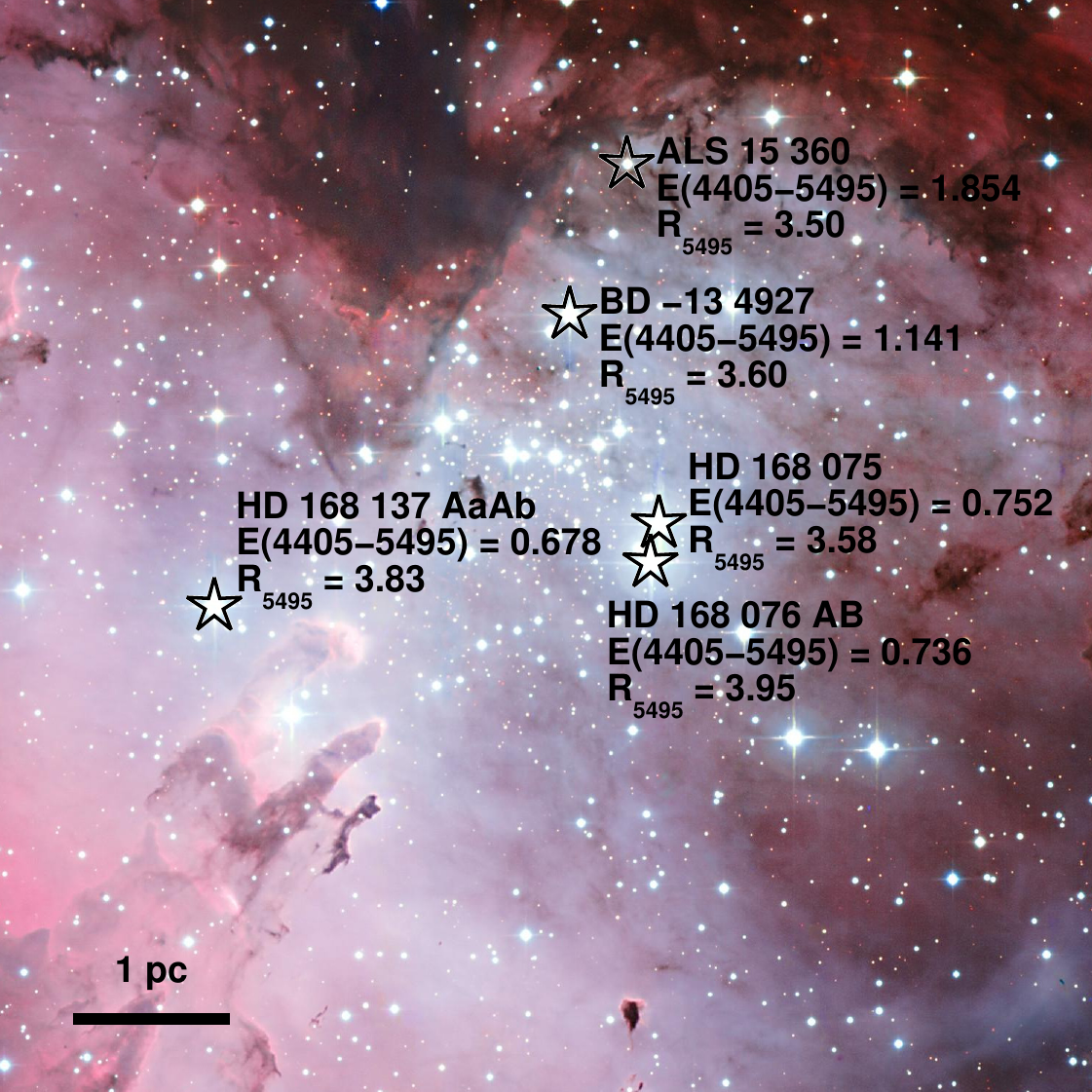}
 }
 \caption{Extinction measurements for O stars in two \HII\ regions: (left) the Carina Nebula and (right) M 16/NGC 6611. The 
          two images are from ESO press releases 1250 ($Bgr$H$\alpha$) and 0926 ($BVR$), respectively. North is towards the 
          top and east towards the left and the approximate physical scale is indicated. Only a section of each \HII\ region
          is shown to better visualize the dust lanes.}
 \label{CarinaM16}
\end{figure}

\begin{figure}[ht!]
 \centerline{
 \includegraphics[width=1.20\linewidth]{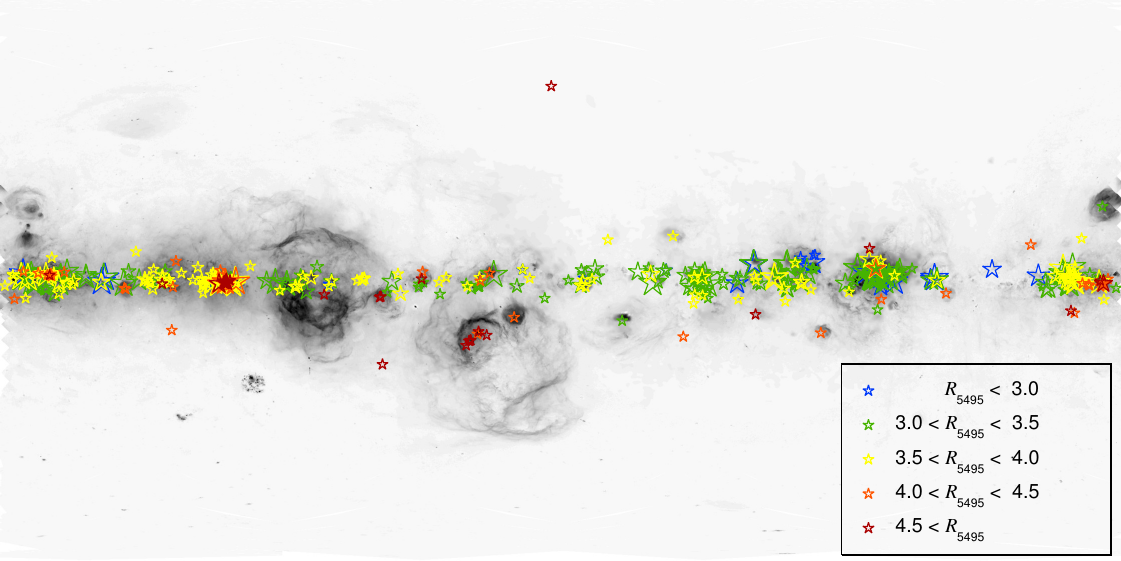}
 }
 \caption{Extinction measurements for Galactic O stars plotted on top of the full-sky H$\alpha$ image of \citet{Fink03}. 
          Different colors are used for five ranges of \RV\ with the size of the symbol increasing with \EBV. The 
          background image is showed in a logarithmic intensity scale and in Galactic coordinates with a Cartesian 
          projection centered on the Galactic anticenter.}
 \label{extinction_map}
\end{figure}

It is important to understand the uncertainties associated with both \EBV\ and \RV. Parameter uncertainties in CHORIZOS are 
calculated by properly sampling the multi-parameter space around the minimum $\chi^2$ location, thus allowing for a correct
characterization of their correlations \citep{Maiz04c}. In general, \EBV\ can be determined reasonably well (uncertainties 
of $\sim 0.02$~mag) at most extinctions as long as there is good-quality photometry. This is easy to understand as the 
intrinsic $(B-V)$ of OB stars $(B-V)_0$ is well determined and, to a first approximation, $\EBV\sim E(B-V)$ (but see 
Fig.~\ref{dEBVRV}) and from Eqn.~\ref{EBV} the uncertainties arise only from those in the measured (Johnson) $B-V$, which 
are of the order of 0.02~mag (see Table~9 in \citealt{Maiz06a} and remember the previous point about realistic external
photometric uncertainties)\footnote{This is just a simplified version of what the code does and assumes that Johnson $BV$ 
photometry is available. In reality, all used magnitudes contribute in some degree to the detrmination of \EBV.}. 
The uncertainties on \RV, on the other hand, are much larger for small values of \EBV\ than for
large ones. The explanation is that \RV\ is a measurement of the type of extinction and in the limit of low extinction 
information about it is limited. Interestingly, the sampling of $\chi^2$ in the multi-parameter space shows that in most
cases the uncertainties in \EBV\ and \RV\ are anticorrelated: this is seen in Fig.~\ref{EBV_RV} by the elongation of the
ellipse uncertainties along the lines of constant \AV. Therefore, the relative uncertainty of \AV\ can be
lower than those of both \EBV\ and \RV, something that is impossible in the absence of an anticorrelation. In other words, 
a method that properly measures extinction combining optical and NIR photometry can arrive at a more precise determination 
of \AV\ (and hence, of $A_V$) than of either \EBV\ or \RV. 

Where does the variability in the extinction law originate? Already \citet{BaadMink37} pointed out that the high values of 
\RV\ in Orion are likely due to a larger than average grain size for the dust particles, a view that has remained
since then. In \citet{MaizBarb18} we reviewed the literature and presented results showing how \HII\ regions have large
internal variability in \RV\ and \EBV, with regions with intense nebular emission typically having high \RV\ and low \EBV\
and regions close to or on top of dust lanes having low \RV\ and high \EBV\ (Fig.~\ref{CarinaM16}). In addition, a map of
the \RV\ measurements in \citet{MaizBarb18} over the whole sky (Fig.~\ref{extinction_map}) shows that high-latitude stars,
which are relatively nearby and for which the Local Bubble contribution should be larger than average, tend to have high
values of \RV. This led us in \citet{MaizBarb18} to propose a scenario in which there are three different ISM regimes
depending on the ambient UV radiation level:

\begin{itemize}
 \item Regions with intense UV radiation have large values of \RV\ ($>4$). This includes \HII\ regions but also cavities 
       filled with hot gas (e.g. the Local Bubble) where UV radiation can travel for long distances.
 \item Regions with low intensity of UV radiation have small values of \RV\ ($<3$). These regions are those with significant
       column densities of CO or those that are easily detected as dust lanes in the optical. 
 \item Regions with intermediate levels of UV radiation have values of \RV\ between 3 and 4. These regions represent a 
       typical warm to cold ISM (excluding \HII\ regions, cavities, and molecular clouds) that fills the majority of the 
       volume in the Galactic disk.
\end{itemize}

Therefore, high levels of UV radiation correspond to large average grain sizes (high \RV). The likely explanation is that 
ISM regions with large values of \RV\ are produced by the selective destruction of smaller dust grains with respect to 
large grains. Such a selective destruction could be caused by thermal sputtering \citep{Drai11} or heating by EUV radiation
\citep{GuhaDrai89,Jone04}.

\subsection{A comparison with \textit{Gaia}~DR3 results}

\begin{figure}[hpt!]
 \vspace{-5mm}
 \centerline{
 \includegraphics[width=0.60\linewidth]{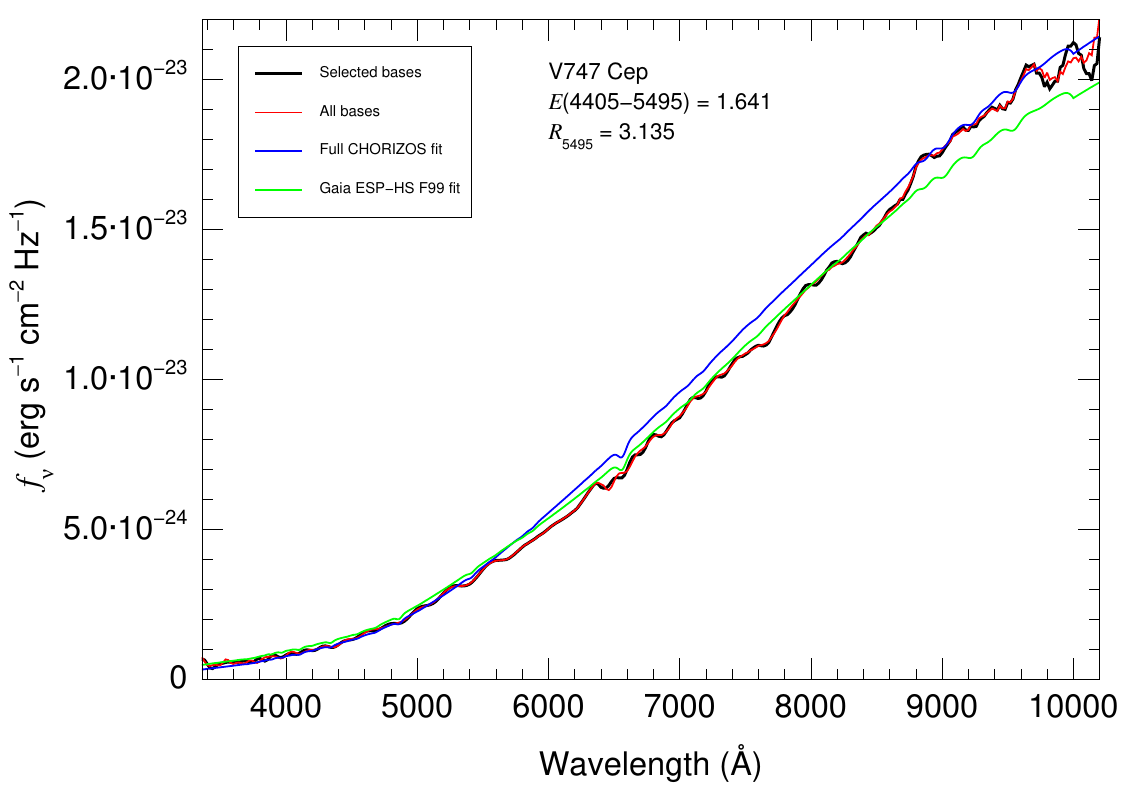}
 }
 \vspace{-1mm}
 \centerline{
 \includegraphics[width=0.60\linewidth]{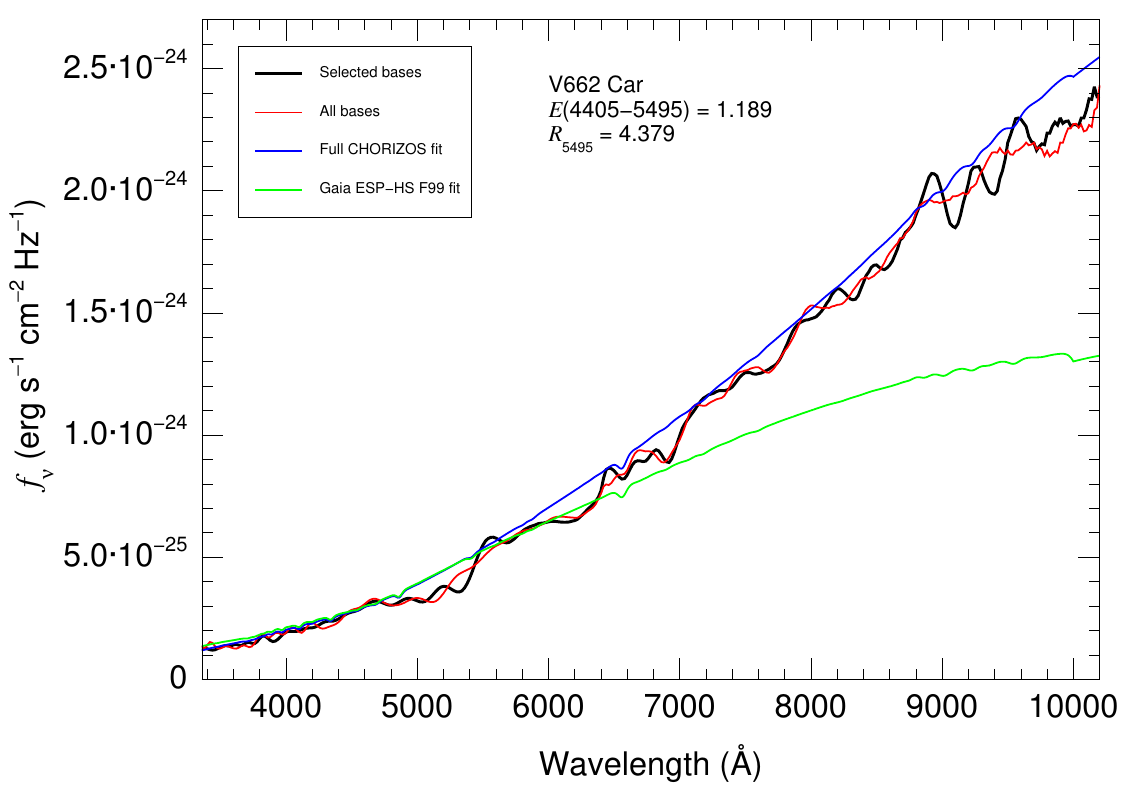} \
 \includegraphics[width=0.60\linewidth]{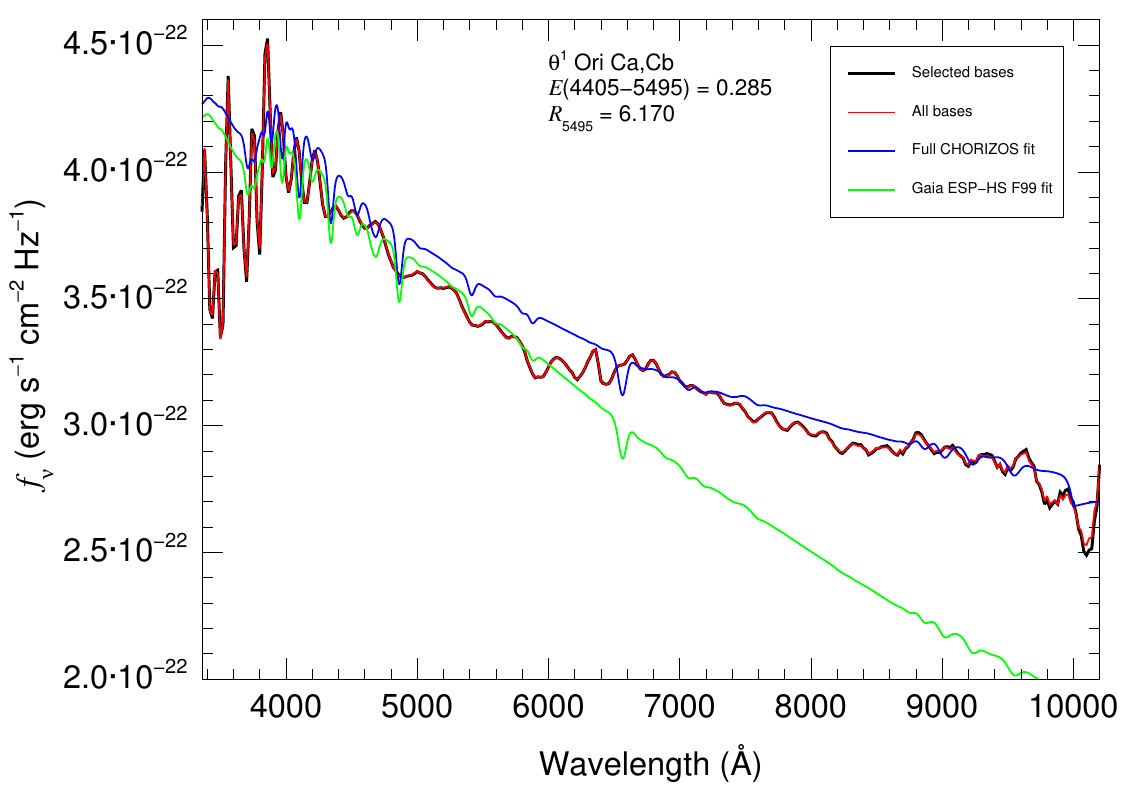}
 }
 \vspace{-3mm}
 \caption{Comparison between the CHORIZOS results of \citet{Maizetal21a} and the \textit{Gaia}~XP results of 
          \citet{Creeetal23a}. The black/red lines are the input XP spectra using two choices of basis functions (to 
          showcase the uncertainty in flux), the blue line is the SED derived from the CHORIZOS fit, and the green line the 
          SED derived from the \citet{Creeetal23a} XP fit (displaced to fit as much of the XP spectra as possible). The 
          names of the stars and their CHORIZOS-derived values of \EBV\ and \RV\ are given in each panel, with the three 
          stars selected to showcase different values of \RV.}
 \label{Gaia_XP}   
\end{figure}

$\,\!$\indent To see the importance of including \RV\ in an analysis of extinction, here I compare the results from
\citet{Maizetal21a} with the \textit{Gaia}~DR3 results of \citet{Creeetal23a}. That paper presents the \textit{Gaia}-DPAC
analysis using the Astrophysical parameters inference system (Apsis). With a variety of information from different
\textit{Gaia} sources, they derive stellar and extinction parameters for a large sample of stars. Here I concentrate on the 
results from the ESP-HS module, the Extended Stellar Parametrizer for Hot Stars, as that is the module which is more
appropriate (and favorable) to use against the OBA-star analysis with CHORIZOS in \citet{Maizetal21a}. 

The ESP-HS module uses a library of hot-star atmospheres to derive \Teff, \logg, $A_0$ (the equivalent to \AV, see above),
and $d$ based on the parallax and the XP spectra (plus the RVS spectra for some stars).
It fixes the equivalent of \RV\ to 3.1 using the family of extinction laws of \citet{Fitz99}. 
As a comparison, the analysis of \citet{Maizetal21a} is based on photometric data (including \textit{Gaia} and 2MASS) and
fits \EBV, \RV, and \logd, fixing \Teff\ and the luminosity class (a proxy for \logg) from the spectral types, and using the
family of extinction laws of \citet{Maizetal14a}. I compare the results from the two procedures for three stars of
different \RV\ in Fig.~\ref{Gaia_XP}.

The predicted CHORIZOS SED is a reasonable fit for the XP spectra of the three stars despite having been obtained prior to
\textit{Gaia}~DR3. There are some small differences, both at small and large wavelength scales, some of which may be due to
small deficiencies in the \citet{Maizetal14a} family of extinction laws and some due to systematic biases in the XP spectra.
Among the first, I note that for V747~Cep there is a hint of the 7700~\AA\ bump and for V662~Car there is a clear
detection. Among the second, many of the wiggles seen are likely artifacts due to the extraction procedure as a series of
basis functions \citep{Carretal21}.

As for the \citet{Creeetal23a} results, the fit is reasonably good for V747~Cep, which has an \RV\ of 3.135$\pm$0.078 that
is compatible with the assumed value of 3.1. On the other hand, the fit is bad for either V662~Car and $\theta^1$~Ori~Ca,Cb,
which have values of \RV\ significantly different from the assumed 3.1. The discrepancy is a natural consequence of the
CHORIZOS fit being a good one, as there is no stellar SED of any \Teff\ with $\RV = 3.1$ that resembles that of an O star
with values of $\RV > 4.0$ and a minimum amount of extinction. In other words, the \citet{Creeetal23a} results are
inconsistent with their own input data because they did not include the real solution among the set of possible ones, an
example of ``fitting a straight line to a parabola'' (see above).

The conclusion of this analysis is that the parameters derived by assuming that \RV\ is 3.1 everywhere are bound to fail in
the cases where \RV\ is significantly different and those cases, at least for OBA stars, are relatively common
(Fig.~\ref{EBV_RV}). As an example, the ESP-HS analysis of V662~Car derives a value of $A_0 = 3.639\pm 0.042$~mag. The
equivalent \AV\ calculated from CHORIZOS is $5.206\pm 0.038$~mag, significantly higher. Another lesson is that NIR data need
to be added to optical (spectro)photometry to correctly fit both the type and amount of extinction: \citet{Creeetal23a}
could not include 2MASS photometry in their analysis due to the DPAC policy of excluding non-\textit{Gaia} data.

\section{Families of extinction laws}

\subsection{History}

$\,\!$\indent As already mentioned, \citet{Cardetal89} or CCM published the first family of extinction laws with a parameter
describing the type of extincion ($R_V$ though it should have been a monochromatic quantity such as \RV) to add to the 
amount of extinction [$E(B-V)$ though it should have been a monochromatic quantity such as \EBV]. It was based on 
$UBVRIJHKL$ photometry for the optical+IR part and IUE spectrophotometry for the UV part, thus covering the range from
1250~\AA\ to 3.5~$\mu$m. As with most 1980s extinction papers, CCM is a UV-centric work, as the access to
that range provided by IUE was the novelty at that time and that was where most of the interesting information was 
available. As it turned out, most papers that ended up using the family of extinction laws eventually applied it to the
optical+IR ranges instead of to the ultraviolet, especially after the first decade\footnote{As proof of the importance of
the paper, it is currently the 11th most cited paper on ADS, with a citation count approaching \num{10000} at the time
of the writing of this contribution.}.

The CCM family is based on 29 low-extinction Galactic sightlines, of which only three have listed values with
$E(B-V) > 1.0$~mag and even for those they are only 1.09~mag, 1.11~mag, and 1.22~mag. The listed covered range in $R_V$ is 
2.6-5.3. The reasons for those limitations are easy to understand: as an UV-centric study, objects with higher reddenings
would have been mostly inaccessible to IUE, and at that time there were few objects with known high values of \RV. For the
three wavelength ranges the following functional forms were used:

\begin{itemize}
 \item \textbf{UV:} An adaptation of the \citet{FitzMass88} functional form (see below).
 \item \textbf{Optical:} A 7th degree polinomial in $x\equiv 1/\lambda$ to interpolate between the central wavelengths of 
       the Johnson photometric bands.
 \item \textbf{IR:} A power law $A(\lambda) \propto \lambda^{-\alpha}$ with a fixed value of $\alpha = 1.61$ from 
       \citet{RiekLebo85}.
\end{itemize}

\begin{figure}[t!]
 \centerline{
 \includegraphics[width=0.60\linewidth]{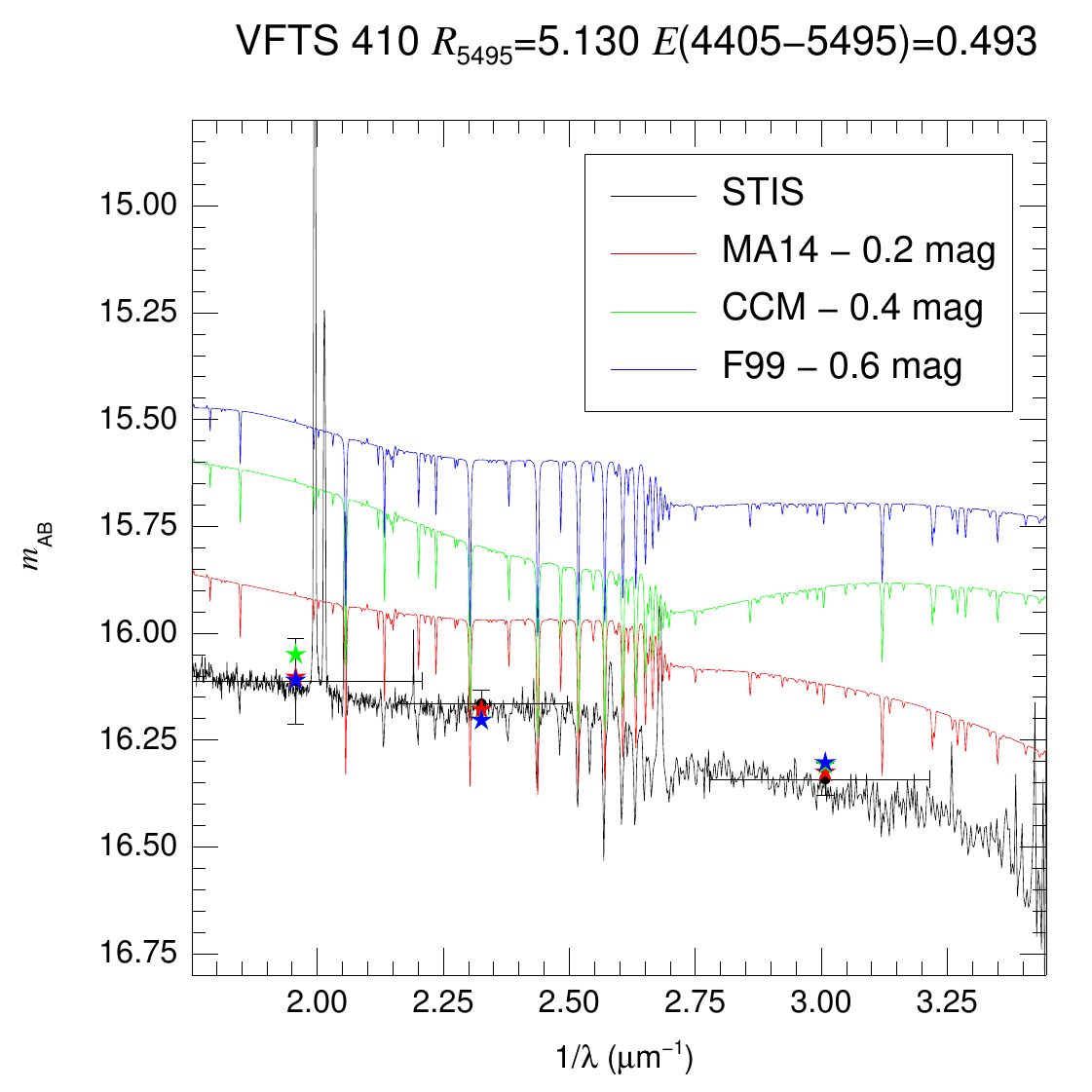} \
 \includegraphics[width=0.60\linewidth]{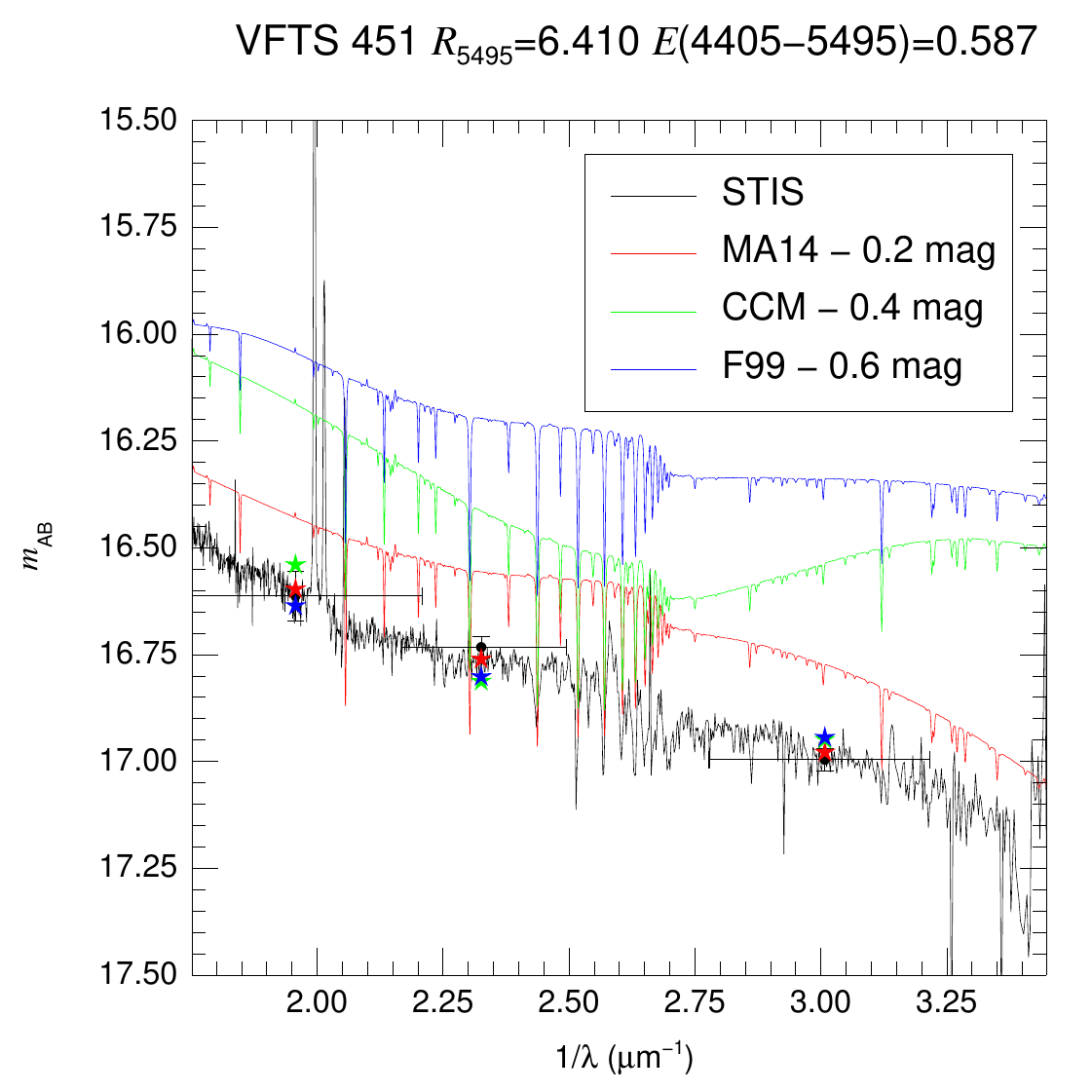}
 }
 \caption{Comparison between the STIS spectrophotometry for two 30 Dor stars in the MA14 sample and the fits using the 
          three families of extinction laws discussed in this contribution. Note how the use of splines in F99
          and MA14 allow them to follow the approximate shape of the spectrophotometry while the seventh degree polynomial
          used by CCM introduces unphysical oscillations in wavelength.}
 \label{oscillations}   
\end{figure}

\begin{figure}[t!]
 \centerline{
 \includegraphics[width=0.60\linewidth]{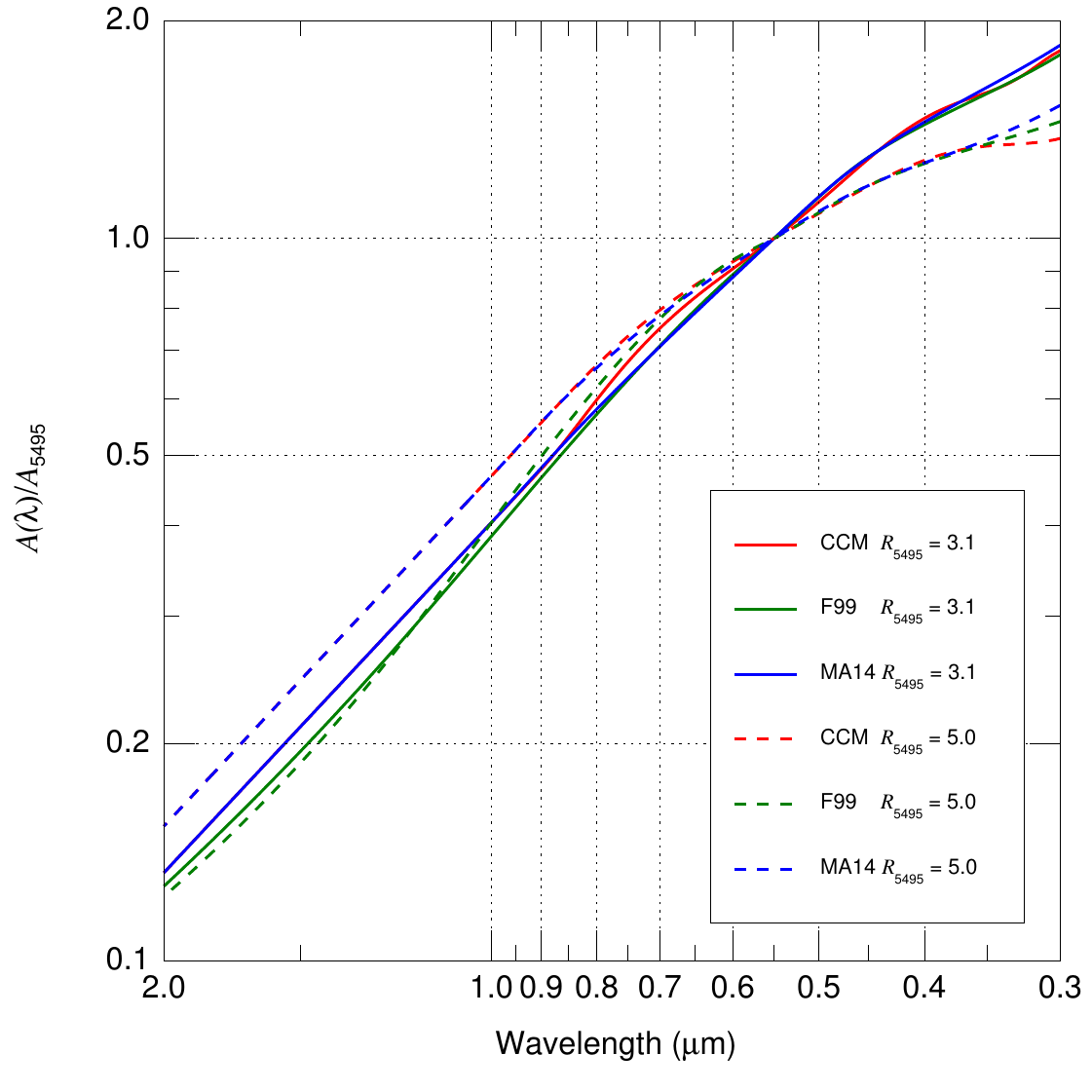} \
 \includegraphics[width=0.60\linewidth]{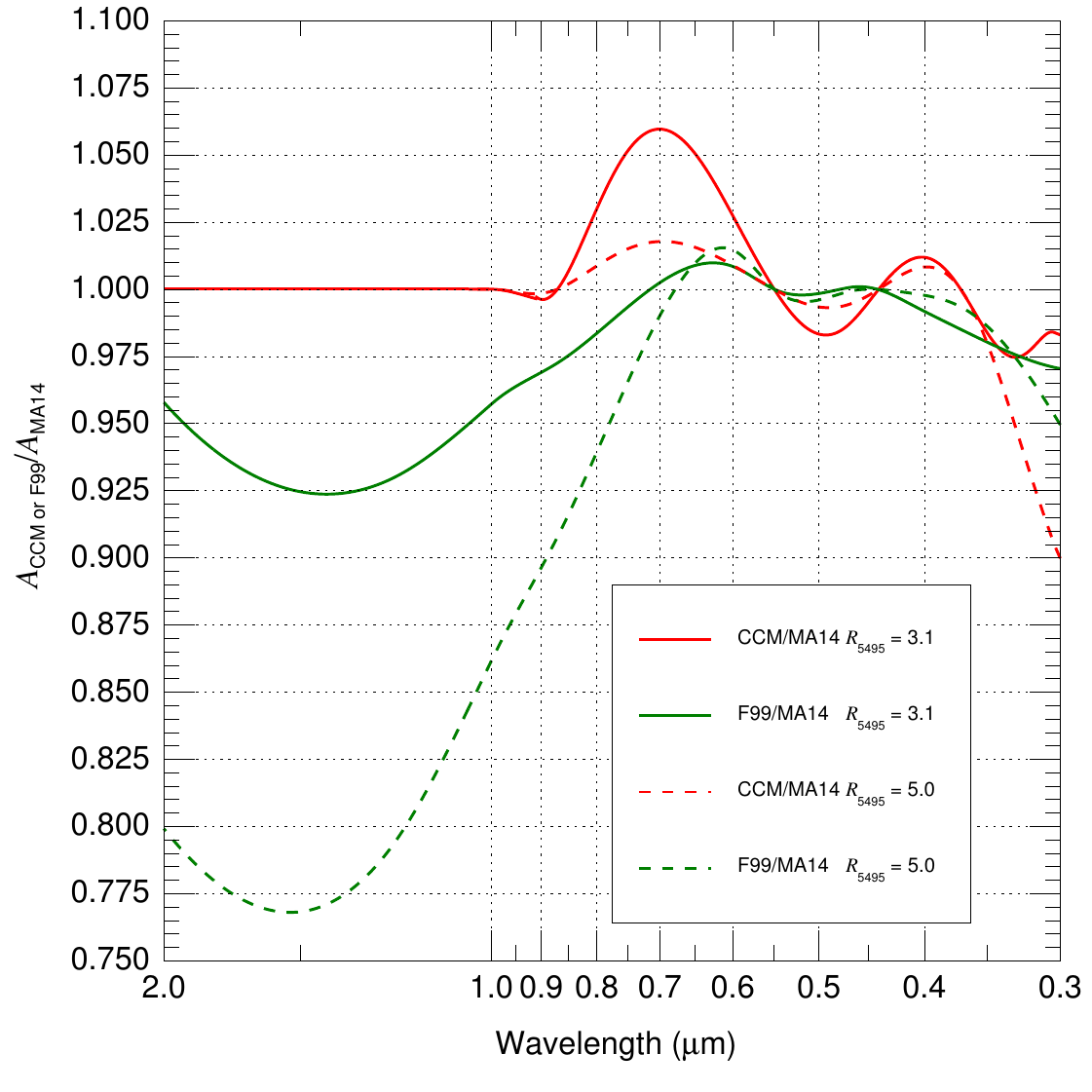}
 }
 \caption{Comparison between the CCM, F99, and MA14 families
          of extinction laws for two values of \RV, 3.0 and 5.1. The left panel shows the extinction laws normalized to 
          the value at 5495~\AA. The right panel shows the ratio of the CCM and F99 laws to the MA14 one of the same \RV.
          Note that the CCM and MA14 laws are identical in the IR for a given \RV.}
 \label{extinction_laws}   
\end{figure}

A decade later, \citet{Fitz99} or F99 produced another family of extinction laws. His sample was larger than that of
CCM but with a similar sampling\footnote{Some of the details of the analysis are hard to follow in that
paper, as the sample is not presented in detail but is a modification of that of \citet{FitzMass90}, their $R_V$ values are
not given, and the notation confuses monochromatic and band-integrated quantities.} in terms of $E(B-V)$, as the paper is
also UV-centric. The main modification with respect to CCM is the use of splines instead of a seventh degree polynomial to 
interpolate between filter central wavelengths and, in that way, avoid unphysical oscillations in wavelength
(Fig.~\ref{oscillations}). The spline interpolation is extended into the IR and, as a result, the functional form does not 
follow a power law there.  Instead, the slope is flatter at wavelengths longer than the $H$ band and steeper at shorter 
wavelengths (Fig.~\ref{extinction_laws}). 

A third family of extinction laws was derived by \citet{Maizetal14a} or MA14 using a sample of 85 OB stars in 30~Dor 
observed with HST/WFC3 $UBVI$ optical and ground-based $JHK$ photometry. The advantages of HST/WFC3 photometry are its 
better calibration and spatial resolution, which leads to a reduction in the systematic effects present in Johnson 
photometry and allows for the detection of possible companions. The covered range in \EBV\ was smaller than those of 
CCM or F99 (an effect offset by the better photometric calibration) but with a better sampling in 
\RV. As opposed to the two previous families, the TLUSTY models used for CHORIZOS were the source for unextinguished SEDs 
instead of real standard stars. 

\begin{figure}[t!]
 \centerline{
 \includegraphics[width=1.20\linewidth]{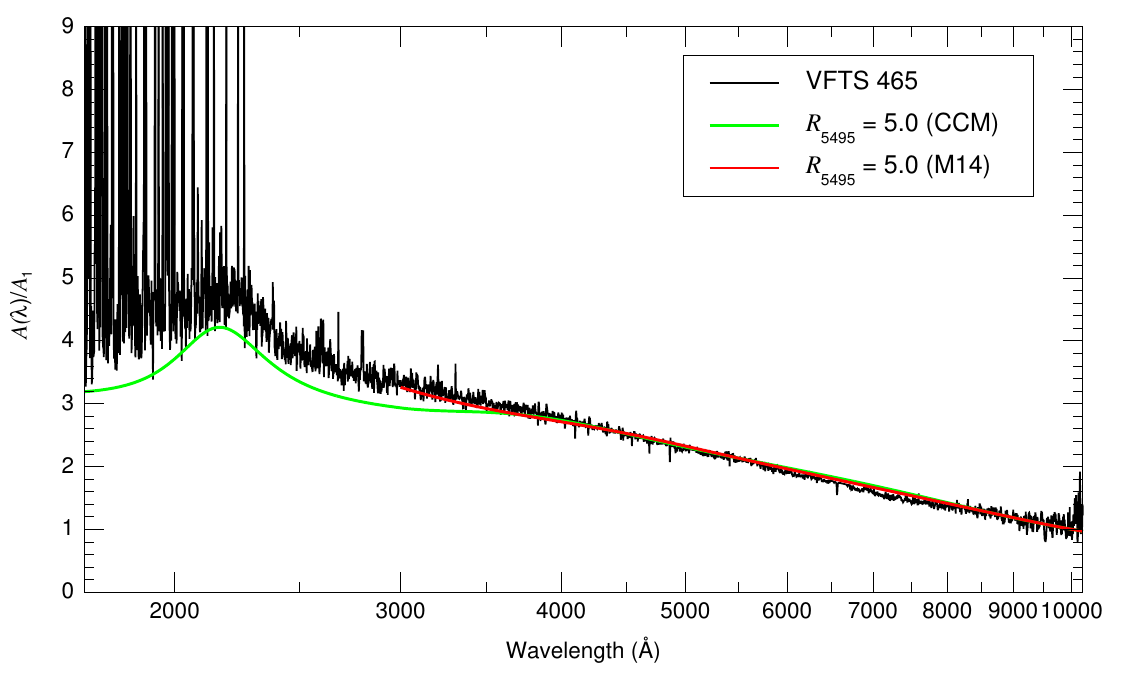} 
 }
 \caption{30~Dor star observed with HST/STIS spectrophotometry and their corresponding SED fits from the photometry in 
          MA14 using the CCM and the new families in that paper. Both fits are very good in the
          optical but in the $U$ band the new laws are significantly better, leading to the possibility that some of the 
          differences in the UV slope arise not from a difference in extinction laws between the Milky Way and 30 Dor but 
          due to a bad stitching between the optical and UV regimes in CCM. I have not removed the effects
          of narrow ISM lines in the NUV, the mismatch in the profiles of stellar lines in the optical, or the noise in the
          FUV and around 1~$\mu$m.}
 \label{VFTS_465_ala1}   
\end{figure}

The idea behind MA14 was to combine the best of both worlds from CCM and F99: the
proven behavior of the first as a function of \RV\ with the use of splines in the optical from the second to eliminate the
unphysical oscillations in wavelength. To that purpose, MA14 started with the CCM family, maintained the IR behavior of 
\citet{RiekLebo85}, and substituted the optical part with a spline equivalent. Then, the photometric residuals were studied 
to discover that the CCM laws were underestimating extinction in the $U$ band (the same is true for F99) 
and corrected for that effect (Fig.~\ref{VFTS_465_ala1}). The shortcomings of the family are that it does not include
the UV; that the IR slope is assumed and not determined (something difficult to do for the range of extinctions of the
sample); and that, being derived from 30 Dor data, its validity for Galactic sightlines was unproven at the time of the
writing of the paper (but see below).

There are other published families of extinction laws but I restrict the analysis to these three, as the first two are the 
ones most often used in the literature and the third one is used for the comparison in the next subsection. In 
Fig.~\ref{extinction_laws} I plot a comparison between the three families, which is a graphical summary of some of the 
points I have made in this subsection.

\begin{itemize}
 \item The CCM and MA14 families follow a power law with $\alpha = 1.61$ and are identical in the NIR, so they are shown as
       superimposed straight lines in the left panel and a value of 1.0 in the right one for that range. The F99 families,
       on the other hand are not power laws in the NIR and they predict less extinction there for the same amount of optical
       extinction. The differences between CCM/MA14 and F99 in the NIR increase with \RV.
 \item The differences between the three families are smaller in the optical than those between CCM/MA14 and F99 in the NIR.
 \item The CCM family has an oscillatory behavior in the optical range due to the use of a seventh degree polynomial in 
       $1/\lambda$ for interpolation. 
 \item Both the CCM and F99 families predict a lower extinction in the $U$ band for the same amount of extinction in the $V$ 
       band.
\end{itemize}

\subsection{The optical/NIR}

\begin{figure}[hpt!]
 \centerline{
 \includegraphics[width=0.60\linewidth]{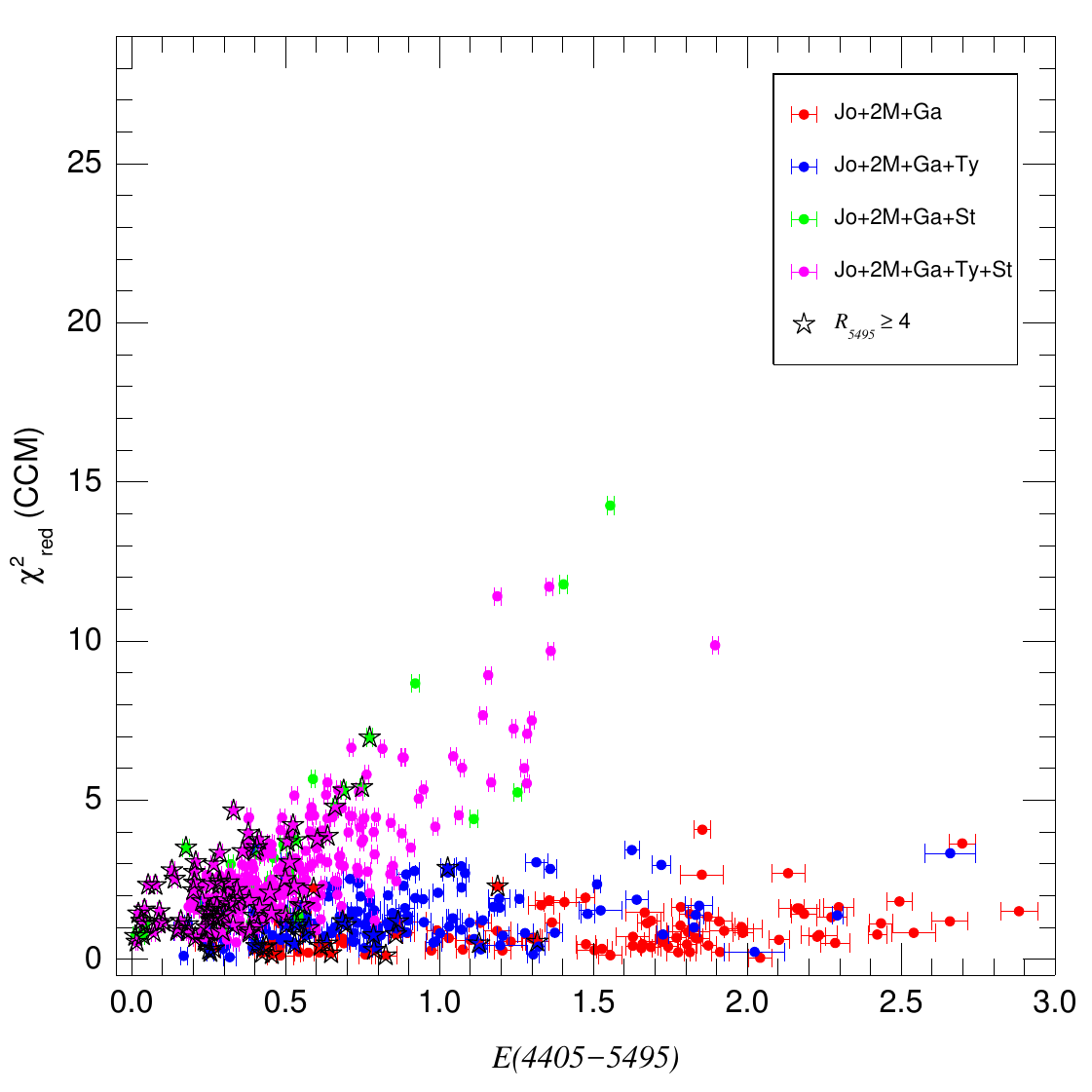} \
 \includegraphics[width=0.60\linewidth]{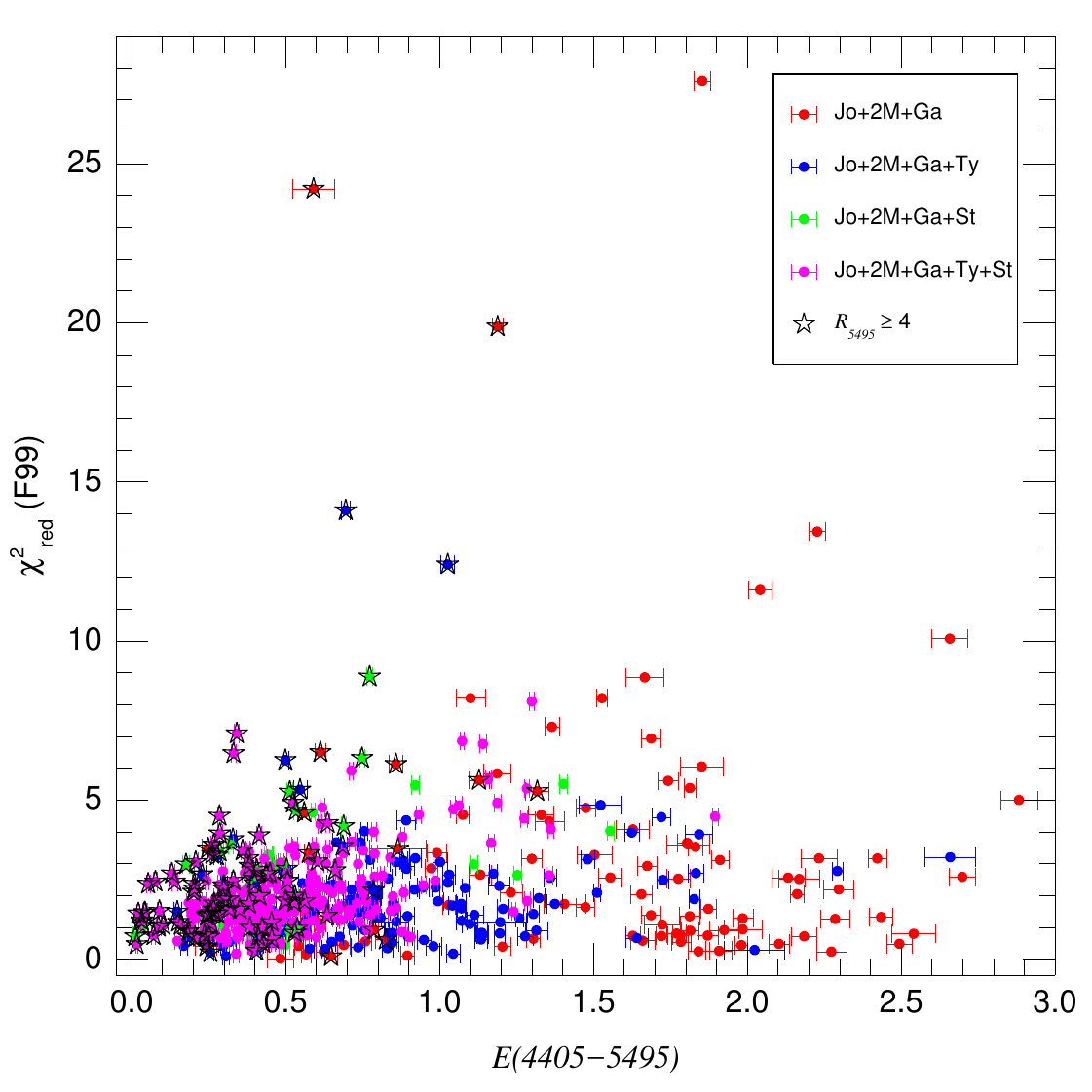}
 }
 \centerline{
 \includegraphics[width=0.60\linewidth]{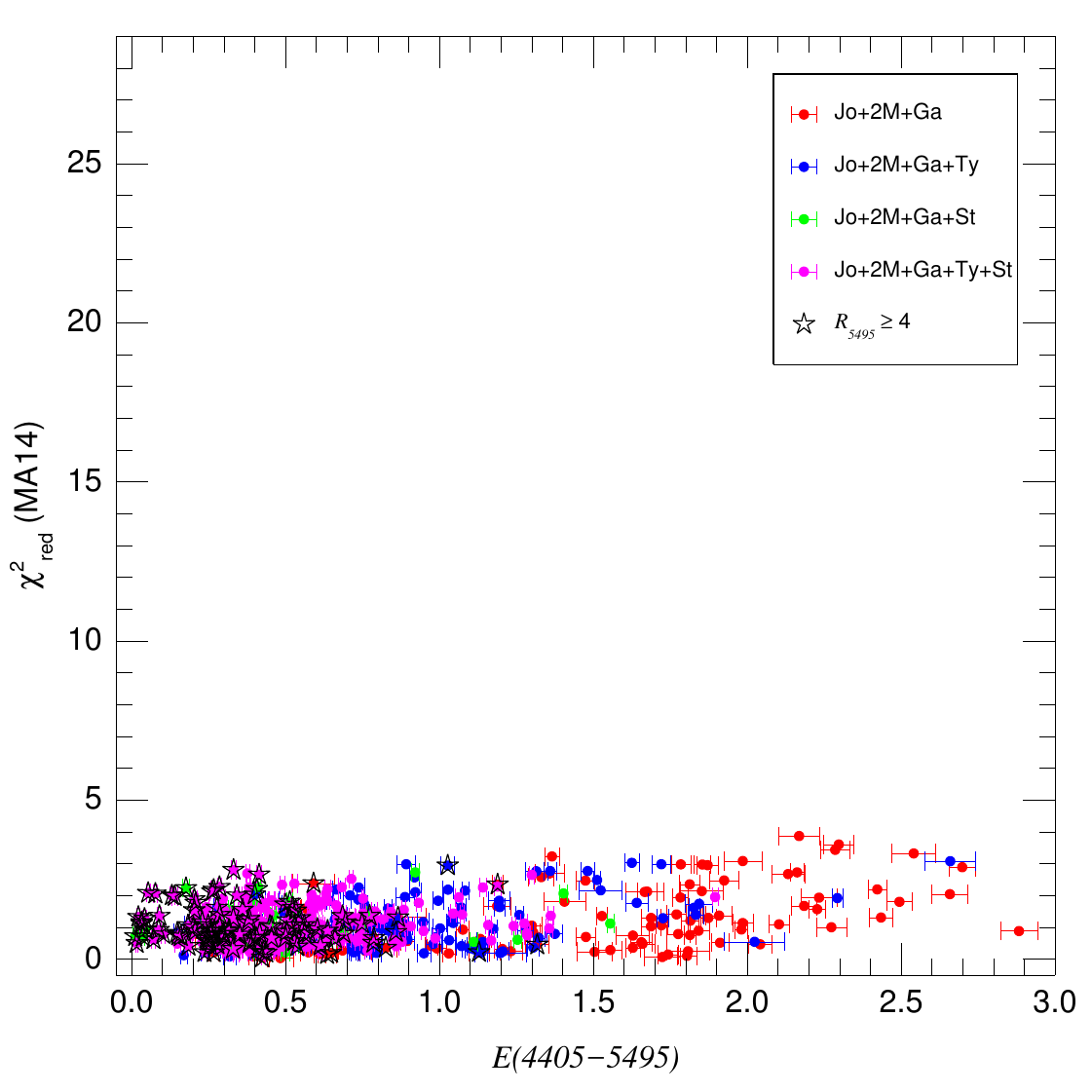} \
 \includegraphics[width=0.60\linewidth]{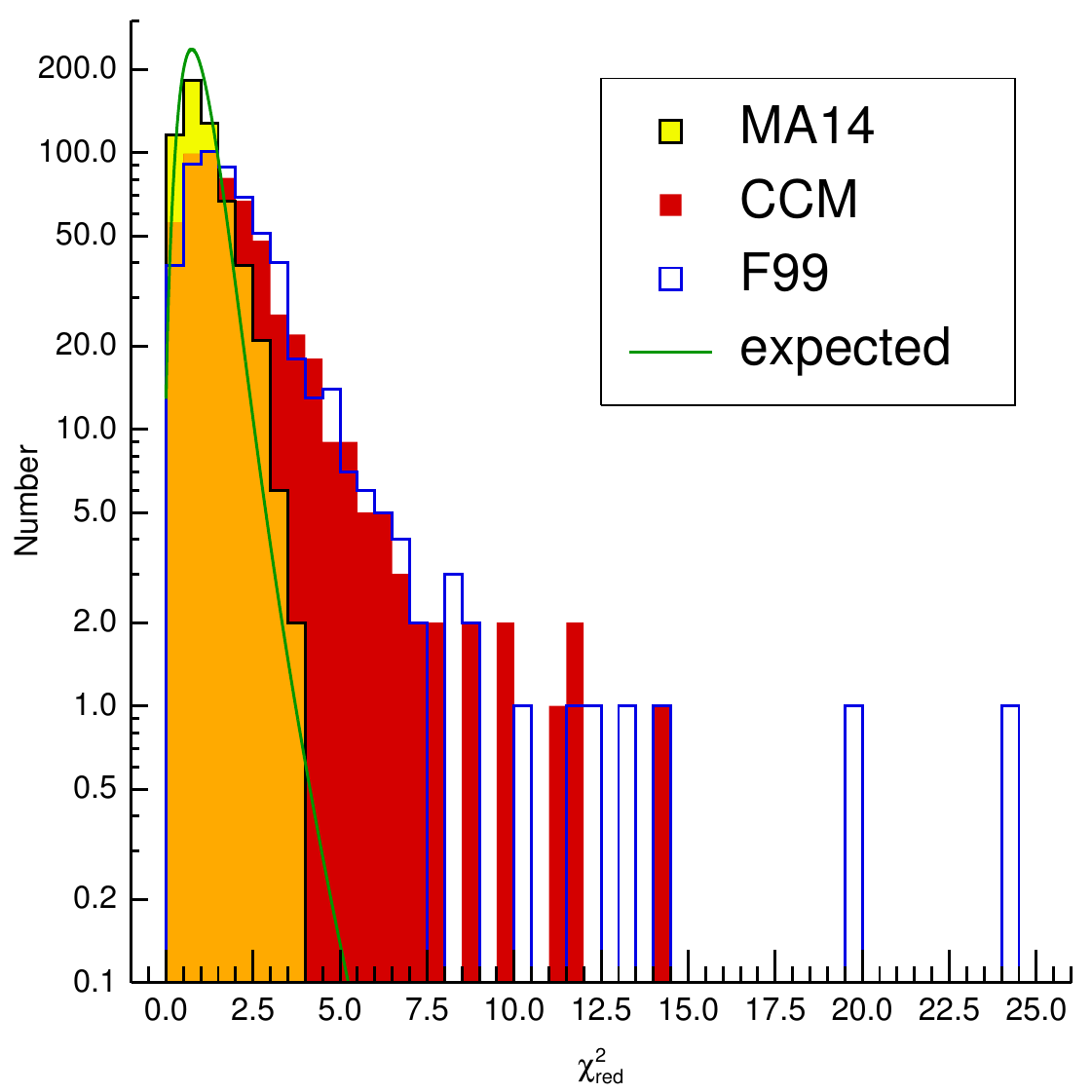}
 }
 \caption{Reduced $\chi^2$ as a function of \EBV\ using the CCM (top left), F99 (top right), and MA14 (bottom left) families
          of extinction laws for the CHORIZOS runs of the Galactic O-star sample of \citet{MaizBarb18}. (bottom right) 
          Reduced $\chi^2$ histograms for the first three plots and expected distribution.}
 \label{comparison}   
\end{figure}

$\,\!$\indent Having presented the characteristics of the three families of extinction laws, the outstanding question is:
which one of the three is a better approximation to reality? The answer was provided by \citet{MaizBarb18}, whose work on
the distribution of \EBV\ vs. \RV\ for 562 O-type Galactic stars was presented in the previous section. That paper also
processed their Johnson+2MASS+\textit{Gaia} photometry (with Tycho-2 and/or Str\"omgren included when available) through
CHORIZOS using the three families. The resulting reduced $\chi^2$ distributions are shown in Fig.~\ref{comparison}

\begin{itemize}
 \item As seen in the bottom right panel, the best results are clearly obtained with the MA14 familiy. The observed reduced
       $\chi^2$ distribution follows the expected one closely, with no target having a value over 4. On the other hand, both
       the CCM and F99 families have extended tails into high reduced $\chi^2$ values that indicate that they cannot provide
       accurate fits for some of the stars in the sample, with four of the five the worst outliers coming from F99. 
       \textbf{The family of extinction laws of \citet{Maizetal14a} provides the best description of Galactic optical 
       extinction for $\EBV < 3.0$ available to date.}
 \item For CCM, the worst results are those where Str\"omgren photometry is included. This is expected given the seventh
       degree polynomial used for that family. The tight correlation shown between the reduced $\chi^2$ and \EBV\ in the
       top left panel for those targets is a telltale sign that the effect is caused by extinction. Other than 
       that, CCM provides a reasonable behavior in the sampled range.
 \item For F99, the worst results are those for high values of \RV, marked with stars in the top right panel. However, there
       is also a significant trend of increasing reduced $\chi^2$ with \EBV\ for targets with near-canonical values of \RV.
 \item I point out once more that the three families of extinction laws were developed using stars that are esentially in
       the leftmost one-third of the \EBV\ range of the CCM/F99/MA14 panels of Fig.~\ref{comparison}. For stars in the 
       remaining two-thirds we are extrapolating the behavior from there. 
\end{itemize}

The analysis of \citet{MaizBarb18} was extended by \citet{Maizetal21a} to include more stars, especially with large values
of extinction (reaching $\EBV = 3.5$). They found that even with the MA14 family, some stars have reduced $\chi^2$ values
larger than 4. This means that the extrapolation from low extinctions is already failing and that an improved family is 
needed. Below I address how new data could be used to achieve that.

\subsection{The UV}

\begin{figure}[t!]
 \centerline{
 \includegraphics[width=1.20\linewidth]{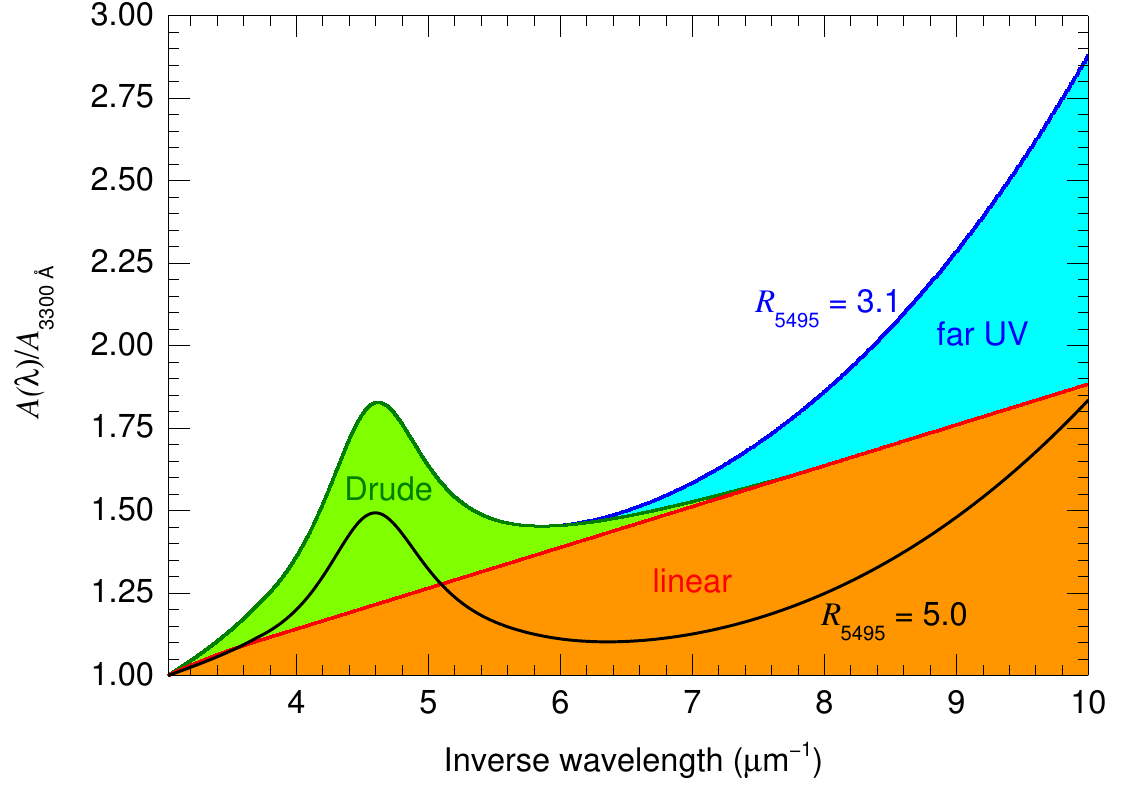}
 }
 \caption{F99 UV extinction for \RV\ of 3.1 and 5.0. The $\RV = 3.1$ case has been decomposed into its three components
          (linear, Drude profile, and far-UV rise).}
 \label{UV_extinction}  
\end{figure}

$\,\!$\indent \citet{FitzMass88} parameterized UV extinction as a function of $x\equiv 1/\lambda$ (in $\mu{\rm m}^{-1}$) as
the sum of three components: a linear term, a Drude function (for the 2175~\AA\  bump) and a far-UV rise. The formula in 
that paper is:

\vfill

\eject

\begin{equation}
\frac{A(\lambda)-A_V}{E(B-V)} = c_1 + c_2 x + c_3 D(x;\gamma,x_0) + c_4 F(x) ,
\label{FM88a}
\end{equation}

\begin{equation}
D(x;\gamma,x_0)  = \frac{x^2}{(x^2-x_0^2)^2 + x^2\gamma^2}, 
\label{FM88b}
\end{equation}

\begin{equation}
F(x) = \left\{\begin{array}{ll}
              0.5392(x-5.9)^2 + 0.05644(x-5.9)^3 & {\rm for} \;\; x \ge 5.9 \\   
              0                                  & {\rm for} \;\; x <   5.9 
              \end{array}\right. .
\label{FM88c}
\end{equation}

As I have previously mentioned, such formulae should not be expressed in terms of band-integrated quantities [$A_V$ or
$E(B-V)$] and in a consistent implementation they should be substituted by their monochromatic equivalents. In
Fig.~\ref{UV_extinction} I represent the UV extinction as implemented by F99 using those formulae but renormalized to 1.0
at the beginning of the UV range and based on \RV. CCM used a similar but slightly different parameterization in the UV but
the overall characteristics are the same. Both CCM and F99 used a number of sightliness to derive relationships that express
the six parameters in Eqns.~\ref{FM88a}-\ref{FM88c} ($c_1$, $c_2$, $c_3$, $c_4$, $x_0$, and $\gamma$) as a function of 
\RV\ (or its equivalent). When combined with the optical/IR results, the final outcome in each paper is a single-parameter
family that covers the three wavelength ranges.

The main effect of increasing \RV\ in either the CCM or F99 families is to make the UV slope flatter (for extreme \RV\
values, the linear component can actually be negative). For the LMC and SMC (see references at the beginning of the previous
section) the most important effect is the reduction in $c_3$, the intensity of the Drude profile. 

As opposed to the results in CCM and F99, the analysis of \citet{FitzMass07} found a very different result: with the
exception of few cases with extreme values of \RV, the UV and IR portions of the Galactic extinction curves are not
correlated with each other. In other words, it is not possible to derive a single-parameter (e.g. \RV) family of extinction
laws for the three wavelength ranges, as UV and IR extinctions behave independently. In a subsequent paper,
\citet{FitzMass09} attempted to derive a two-parameter family of extinction laws but, as shown in Appendix E of
\citet{MaizBarb18}, the analysis was flawed because they inadvertantly made one parameter a function of the other one. In a
recent paper involving those authors \citep{Gordetal23} they derive a new family of extinction laws based on a single
parameter for the UV/optical/IR ranges and only reference \citet{FitzMass07} in passing.

\subsection{The IR}

\begin{figure}[t!]
 \centerline{
 \includegraphics[width=1.20\linewidth]{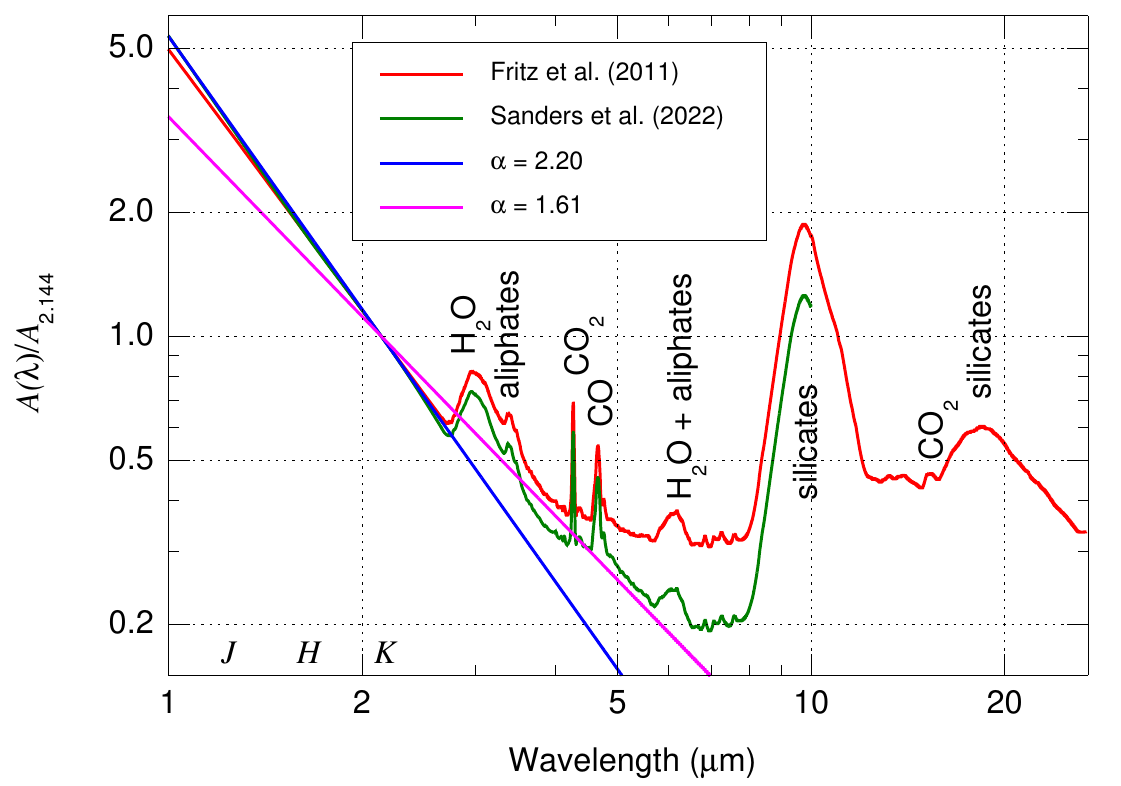}
 }
 \caption{IR extinction laws normalized to 2.144~$\mu$m from \citet{Fritetal11} and \citet{Sandetal22}. As a comparison, 
          two power laws with $\alpha$ of 2.20 and 1.61 are also shown. The location of the $JHK$ bands and the
          identification of the main absorption features are printed.}
 \label{IR_extinction}  
\end{figure}

$\,\!$\indent As already mentioned, \citet{RiekLebo85} approximated the NIR extinction law with a power law:

\begin{equation}
A(\lambda) = A_1 \lambda^{-\alpha},
\end{equation}

\noindent where $A_1$ is the extinction at 1~$\mu$m, $\alpha$ is the power-law exponent, and $\lambda$ is expressed in
$\mu$m. \citet{RiekLebo85} obtained $\alpha = 1.61$, which is the value that CCM and MA14 used. On the other hand, most 
results of the last 15 years 
\citep{SteaHoar09,Fritetal11,Schuetal15,Damietal16b,Alonetal17a,Noguetal18a,Maizetal20a,Sandetal22} have 
obtained significantly higher values of $\alpha$ between 2.11 and 2.47, with most in the range $2.2\pm0.1$. One of 
the reasons for the discrepancy with the \citet{RiekLebo85} result is that the latter inadvertantly included Cyg~OB2-12 in 
their sample, a B-type variable hypergiant with a strong wind and a significant IR excess \citep{Salaetal15,Nazeetal19} 
which flattens the derived extinction law if not corrected for. Also, most recent papers show little sightline-dependence 
variability either with direction or amount of extinction \citep{SteaHoar09,WangJian14,Schuetal15,Noguetal18a,Maizetal20a} 
and a progressive flattening of the slope as we move from the $JH$ bands to the $K$ band and beyond 
\citep{Fritetal11,Noguetal19,Sandetal22}, indicating that IR extinction cannot be approximated by a power law over the 
whole range \citep{Hoseetal18}.

There are no known strong absorption features in the $JHK$ bands (or NIR) but the situation changes in the MIR, with
significant absorption features arising from H$_2$O, CO, CO$_2$, aliphates, and silicates
\citep{Fritetal11,Sandetal22,Decletal22}, see Fig.~\ref{IR_extinction}. The flattening detected by some authors between the
$JH$ and $HK$ bands becomes more significant in the 3-8~$\mu$m region (once the absorption features are removed) with 
values of $\alpha \sim 1.5$ or even lower.

One apparently discrepant voice is that of \citet{Decletal22} who measure $\alpha = 1.7$ using relatively low-extinction OB 
stars. The difference is not due to the nature of the targets (as OB stars inhabit, in principle, different typical 
environments than late-type giants, see \citealt{MaizBarb18} for the effects on extinction), as \citet{Damietal16b}
measured a value of $2.13\pm 0.08$ also using OB stars (albeit more extinguished). Instead, I suspect the difference is
caused by a combination of two effects:

\begin{itemize}
 \item The low extinction of their targets (compared to most other similar studies) makes their results more subject to
       systematic errors. Indeed, the values they obtain for $\alpha$ vary between 1.36 and 2.20, a range much larger than 
       what is seen in other studies. Also, their Fig.~5 reveals that the power-law fit is not a good approximation for the
       data of some of their wavelength regions and stars.
 \item Their wavelength range is large, covering 0.8-5.5~$\mu$m for some of their stars, as opposed to the narrower $JHK$
       range of most studies above. We have already seen that the extinction law flattens for wavelengths longer than the
       $K$ band and the same is true in the 0.8-1.0~$\mu$m region (Fig.~\ref{extinction_laws}). Therefore, an average 
       $\alpha$ in the 0.8-5.5~$\mu$m range is likely to be lower than its equivalent for the $JHK$ bands. 
\end{itemize}

\section{The future}                    

$\,\!$\indent Extinction makes stars look faint but the future for extinction studies looks bright. The most significant
improvement will be spectrophotometry substituting the role of photometry, thus dissipating the doubts over the 
functional form of the extinction law and allowing for the detection of possible structures at intermediate wavelength 
scales, such as the already known 7700~\AA\ bump. For the case of the Magellanic Clouds (and, later on, for other galaxies)
there is plenty of room to significantly increase the number of studied sightlines. Regarding spectrophotometry, 
in the optical \textit{Gaia}~XP is likely to play the major role thanks to its large sample size and dynamic range and 
despite its poor spectral resolution. That may be complemented with HST/STIS data for a few sightlines, especially if the 
mission does not end soon. In the IR there is still ISO and Spitzer archival data to be fully
analyzed but new data may come from ground-based observations and from JWST, though in the latter case I am somewhat
pessimistic that panels will assign much time to bright objects (as many extinguished stars are in the IR). In the UV HST
is likely to still be the main contributor in the short term with projects such as ULLYSES \citep{Romaetal23} until new
missions are launched.

Another line that should be explored is the improvement in calibrations provided by \textit{Gaia} (spectro)photometry 
applied to ground-based photometric surveys. That will allow detailed extinction studies to complement \textit{Gaia} with
narrow and intermediate bands. An example of that type of study is GALANTE \citep{Maizetal21d}

Extinction should be understood as the effect of ISM material on astronomical light sources, independently of
whether the material is ``generic'' dust extinguishing the continuum or is in lines/bands assigned to a known or suspected
material. This is already done in the NIR (Fig.~\ref{IR_extinction}) and in the UV with the 2175~\AA\ band (long suspected
to originate in graphite). Therefore, the extinction law should eventually incorporate atomic and molecular lines, which
would introduce further complications. First, is that different environments in terms of composition, depletion, and
ionization state will contribute to different degrees to extinction, thus necessitating multi-parameter families. Second, 
some lines like the optical Na\,{\sc i} and Ca\,{\sc ii} doublets easily saturate near the Galactic plane and their 
measured EWs depend not only on the column density but on the temperature and kinematics along the sightline. However, even 
approximately incorporating them into an extinction line would provide us information on the conditions associated with 
different components of the ISM, such as the excess Ca\,{\sc ii} absorption associated with high-velocity clouds
\citep{RoutSpit51,RoutSpit52}.

The final components that should be added to the extinction law are DIBs, ISM bands discovered a century ago 
\citep{Hege22,Merr34} but of still uncertain nature save a few adscribed to fullerenes \citep{FoinEhre94}. It is known that
there are different types of sightlines \citep{Kreletal97,Camietal97}, named $\sigma$ and $\zeta$ after the prototypes 
$\sigma$~Sco and $\zeta$~Oph, that sample different regimes of the ISM based on the exposure to UV radiation. The DIB EWs 
are well correlated with \EBV\ and the sightline type is weakly correlated with \RV\ \citep{Maiz15a}, so there is hope that 
a multi-parameter extinction law could incorporate a significant part of the DIB behavior.

\section*{Acknowledgements} 
I would like to thank the LOC for organizing a nice meeting in such a beautiful location.
I acknowledge support from the Spanish Government Ministerio de Ciencia, Innovaci\'on y Universidades, Agencia Estatal de 
Investigaci\'on (\num{10.13039}/ \num{501100011033}), and FEDER/UE through grant PGC2018-0\num{95049}-B-C22 and grant
PID2022-\num{136640}NB-C22 from the Consejo Superior de Investigaciones Cient\'ificas (CSIC) through grant 2022-AEP~005. 

\newpage

\bibliographystyle{aa}
\bibliography{general}

\end{document}